\pdfoutput=1

\documentclass[12pt,a4paper]{article}
\usepackage[T1]{fontenc}
\usepackage[a4paper,margin=2.5cm]{geometry}
\usepackage{amsmath,amssymb}
\usepackage{graphicx}
\usepackage{booktabs}
\usepackage{array}
\usepackage{float}
\usepackage{caption}
\newcolumntype{L}[1]{>{\raggedright\arraybackslash}p{#1}}
\newcolumntype{R}[1]{>{\raggedleft\arraybackslash}p{#1}}
\usepackage[round,authoryear]{natbib}
\usepackage[hidelinks]{hyperref}

\usepackage{setspace}
\bibpunct{(}{)}{;}{a}{}{,}
\providecommand{\doi}[1]{doi:\,\texttt{#1}}

\title{\bfseries Calibrated uncertainty for wide-angle crustal models:
how firmly is the Dharwar Craton Moho actually constrained?}

\author{Deepak Kumar$^{1,2}$\thanks{Corresponding author:
\texttt{deepak.kumar@igf.edu.pl}} \and Laxmidhar Behera$^{2}$
\and Wojciech Czuba$^{1}$}

\date{%
\small
$^{1}$Institute of Geophysics, Polish Academy of Sciences,
Ksi\k{e}cia Janusza 64, 01-452 Warsaw, Poland\\
$^{2}$CSIR-National Geophysical Research Institute, Uppal Road,
Hyderabad 500007, India}

\begin{document}
\maketitle

\begin{abstract}
\noindent
Depth uncertainties for wide-angle crustal models are usually obtained by
propagating an assumed picking error, yet picks from one shot gather are
mutually dependent and velocity--depth trade-offs act at the level of the
model class. We revisit the 210-km Perur--Chikmagalur profile across the
Dharwar Craton using a permanently identified ledger of 2,000 primary
$P_1$--$P_2$ first arrivals and 9,919 reflections from seven shots, with 852
later-turning $P_4$ observations held aside for a conditional test. At fixed
production velocity, a reciprocity-based fast-marching table makes it
affordable to refit all six reflectors inside nested complete-shot validation
and a receiver-coordinate block bootstrap. Withheld reflections are predicted
to 68~ms RMS with standardized mean-squared residual 2.07 and 85 per cent
coverage inside nominal 95 per cent limits; calibration is strongly
phase-dependent, the Moho reflection specifically giving 73~ms and 2.12.
First arrivals degrade from a 17-ms all-data RMS to 103~ms on untouched
shots, driven by aperture extrapolation and a parameterisation blind to the
weathered near-surface. Exact window means place the Moho at 46.9~km beneath
the western block and 39.4~km beneath the central block, a raw contrast of
7.47~km. Resampling receivers within shots gives a $\pm0.10$-km conditional
spread; deleting whole shots gives $\pm0.40$~km, and two independent
experiments agree the raw contrast is biased low by 0.36--0.48~km.
Admissible velocity models span 6.46--8.64~km. Paired LVL-boundary
reflections at 1,260 receivers separate beyond uncertainty everywhere, though
the internal velocity shape stays unresolved. Sampling uncertainty and
model-class uncertainty must be reported apart.
\end{abstract}

\medskip
\noindent\textbf{Keywords:} wide-angle seismics; traveltime tomography;
uncertainty calibration; Moho; Dharwar Craton; low-velocity layer.

\section{Introduction}

Measurements of continental crustal thickness still rest largely on
controlled-source wide-angle profiling. \citet{spence1985} and
\citet{luetgert1990} established much of the recording practice,
\citet{holbrook1992} synthesised the velocity structure such experiments
recover, and later surveys by \citet{carbonell2000} and \citet{snyder2009}
extended the approach to whole-lithosphere targets. Models built this way
underpin what we believe about crustal growth, isostatic balance and how deep
surface structures reach. Their quoted uncertainties, however, are almost
always inherited from whatever error was assigned to the traveltime picks.
\citet{mcmechan1980} and \citet{huang1986} set out the propagation logic early,
and it was carried into routine practice by the ray-trace inversion of
\citet{zelt1992}; a model that satisfies the picks to within their stated error
is pronounced adequate, and the depth it returns is quoted with an uncertainty
obtained by pushing that same stated error through the inversion, a step whose
limits \citet{zelt1999} examined directly.

There is a specific flaw in this reasoning. Errors made while picking a shot
gather are not independent of one another. A mistimed shot instant, a wrong
near-surface correction, or a phase followed one cycle late will displace every
trace in that gather in the same direction. Consequently the number of
genuinely independent observations is very much smaller than the number of
picks, and an uncertainty computed as though each pick were independent is too
small by a factor the calculation itself has no way of exposing. The
consequence is easy to recognise in the literature on this very craton, where
\citet{reddy2000}, \citet{devi2001}, \citet{kumar2003}, \citet{borah2014} and
\citet{beherakumar2022} report crustal thicknesses differing by several
kilometres while each attaches an uncertainty of a few hundred metres.

We proceed differently. Instead of propagating an assumed error, we measure
predictive error directly: complete shot gathers are withheld, the model is
refitted on what remains, and the untouched gather is then predicted. Grouping
at the shot level is what stops traces from a single acquisition gather
appearing in training and test sets at once, a precaution whose necessity
\citet{varma2006} and \citet{cawley2010} demonstrated for nested selection and
\citet{roberts2017} for spatially structured data. The remaining axes are then
kept apart from one another: spatial dependence inside a gather is probed by a
receiver-coordinate block bootstrap, the leverage of individual shots by
delete-one-shot refits, velocity--depth ambiguity by a nonlinear perturbation
ensemble, and regularisation together with acquisition bias by targeted
synthetic recovery. At no point is any one of these diagnostics relabelled as a
single distribution-free uncertainty; in particular, with only seven clusters
available we make no appeal to the conformal machinery of \citet{vovk2005}.

Our subject is the Perur--Chikmagalur profile, a 210-km line trending
north-east to south-west with seven shot points, recorded on three components
by CSIR-NGRI. Following \citet{radhakrishna1986}, the Dharwar Craton is read as
an Archean nucleus split by the Chitradurga Shear Zone, a division supported
seismologically by \citet{singh2004} and \citet{borah2014}; because the profile
crosses that boundary, the thickness contrast between the flanking blocks is
the obvious target. A second and considerably more demanding test is available
in the same data, which carry a clear expression of an upper-crustal
low-velocity layer (LVL); \citet{kumarbehera2023} mapped this feature along the
same line, interpreting an eastward-dipping low-velocity interval that
sandwiches the Dharwar schist belts within the upper crust, and the present
study asks how firmly such an interval is actually constrained.
No ray turns inside an LVL, but a later branch may
cross it and reflections may return from both of its boundaries, so the
constraint has to be assigned phase by phase instead of being inferred from the
first-arrival subset alone. This is the classical hidden-layer difficulty
discussed by \citet{greenhalgh1977}, and it is why \citet{nielsen2000} argued
for combining independent data types when wide-angle geometry alone is
ambiguous. Low-velocity zones are reported at several lithospheric levels,
including the upper mantle beneath this craton \citep{mall2012} and beneath
continents generally \citep{thybo2006}, and their detection is correspondingly
delicate.

Our aims, in sequence, are to establish which phase assignments and model
components can be defended; to test the assigned picking uncertainties against
withheld shots; to hold receiver dependence, shot leverage, recovery bias and
velocity--depth ambiguity apart from each other; and to determine what the two
LVL-boundary reflections and the separate $P_4$ branch actually require. We
place the emphasis throughout on estimands and uncertainty statements that can
be regenerated from permanent observation identifiers.

\section{Data}

\subsection{Acquisition and picks}

\begin{figure}[H]
\centering
\includegraphics[width=0.96\textwidth,height=0.84\textheight,keepaspectratio]{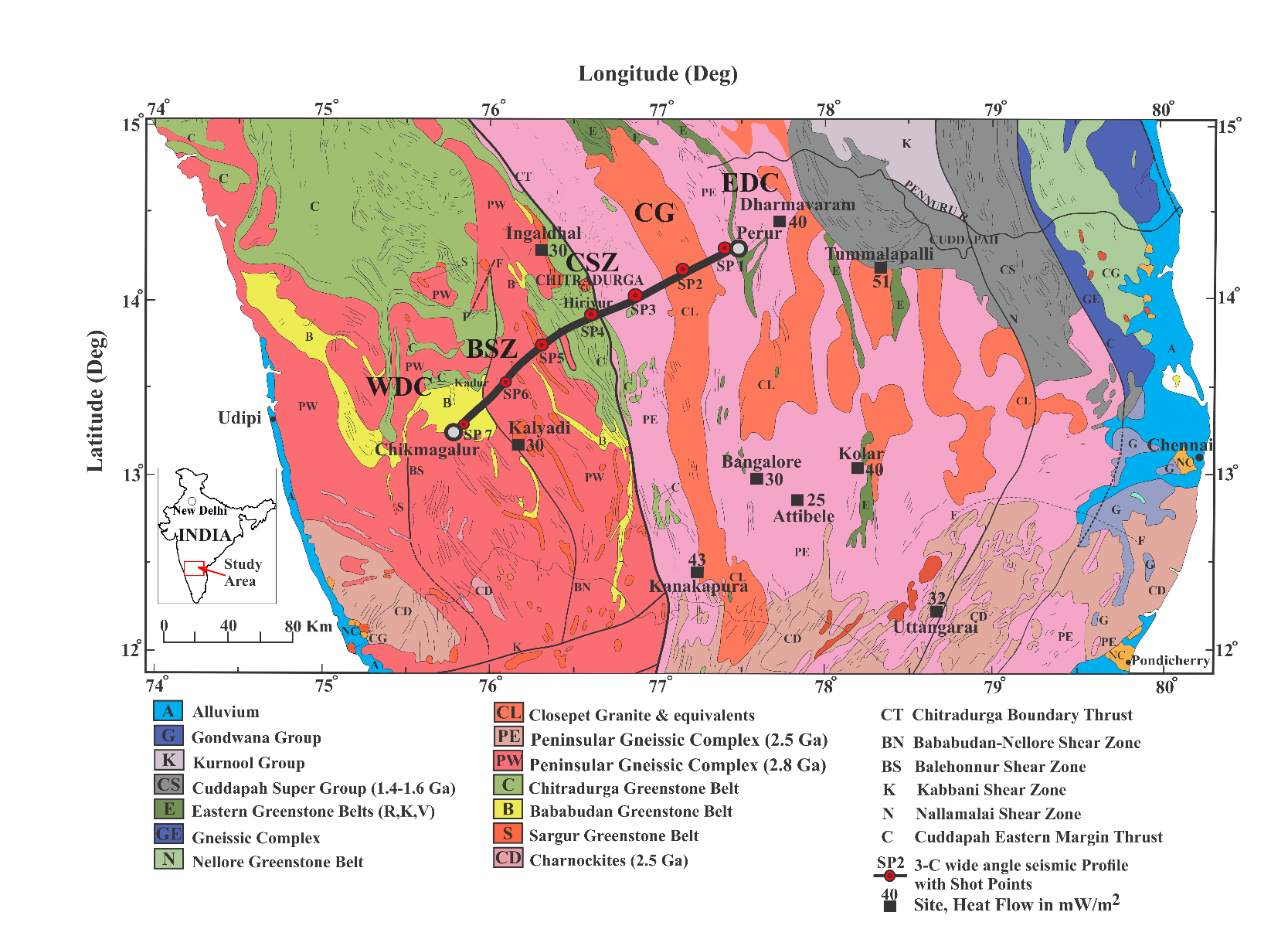}
\caption{Geological setting of the Dharwar Craton and location of the
Perur--Chikmagalur three-component wide-angle seismic profile, adapted from
\citet{kumar2022thesis} and \citet{beherakumar2022}, with the geological base
modified there after GSI and ISRO (1994). The 210-km line runs from Chikmagalur
in the western Dharwar Craton (WDC) to Perur in the north-eastern block, which
we label CDC throughout, and SP1--SP7 are marked. CSZ, Chitradurga Shear Zone;
BSZ, Balehonnur Shear Zone; CG, Closepet Granite. This inherited map supplies
geographical context and is not a product of the present inversion.}
\label{fig:map}
\end{figure}

The line (Fig.~\ref{fig:map}) extends 210~km from Chikmagalur in the south-west
to Perur in the north-east on a bearing near N55$^\circ$E, with seven shot
points at 30--36~km intervals and receivers along its length. Throughout this
paper distance increases towards the north-east in the model coordinate, so
$x=0$ falls at Chikmagalur in the western Dharwar Craton (WDC) and
$x=210$~km at Perur. Because shots are numbered from the north-eastern end,
SP1 sits at $x=199.73$~km and SP7 at $x=0.10$~km.

Two generations of the original \textsc{rayinvr} pick files had to be
reconciled, the later deep-crustal file having dropped the two shallow
reflected branches. Between them they hold 12,918 records: seven shot-location
markers and 12,911 phase observations, reaching source--receiver offsets of
175.1~km and traveltimes of 28.4~s. Every record carries a permanent
identifier assembled from its source-file generation, shot, receiver
coordinate, original code and source-line number. Of these observations, 2,000
unique $P_1$--$P_2$ first arrivals and 9,919 unique reflections enter the
primary inversions, and no identifier appears in both families. A further 852
$P_4$ arrivals are reserved for the conditional later-branch LVL analysis. The
remaining records are set aside: 28 $P_1$ observations lying within 1~km of a
source cannot be used with the finite-radius fast-marching operator, and 112
near-coincident $P^2/P_2$ records are flagged as observationally ambiguous and
withheld from the primary $P^2$ set. These five categories exhaust the ledger,
since $2{,}000+852+9{,}919+112+28=12{,}911$.

\begin{figure}[H]
\centering
\textbf{(a) Shallow SP2 observed gather and interpreted picks}\par
\includegraphics[width=0.96\textwidth,trim=0 340bp 0 0,clip]
  {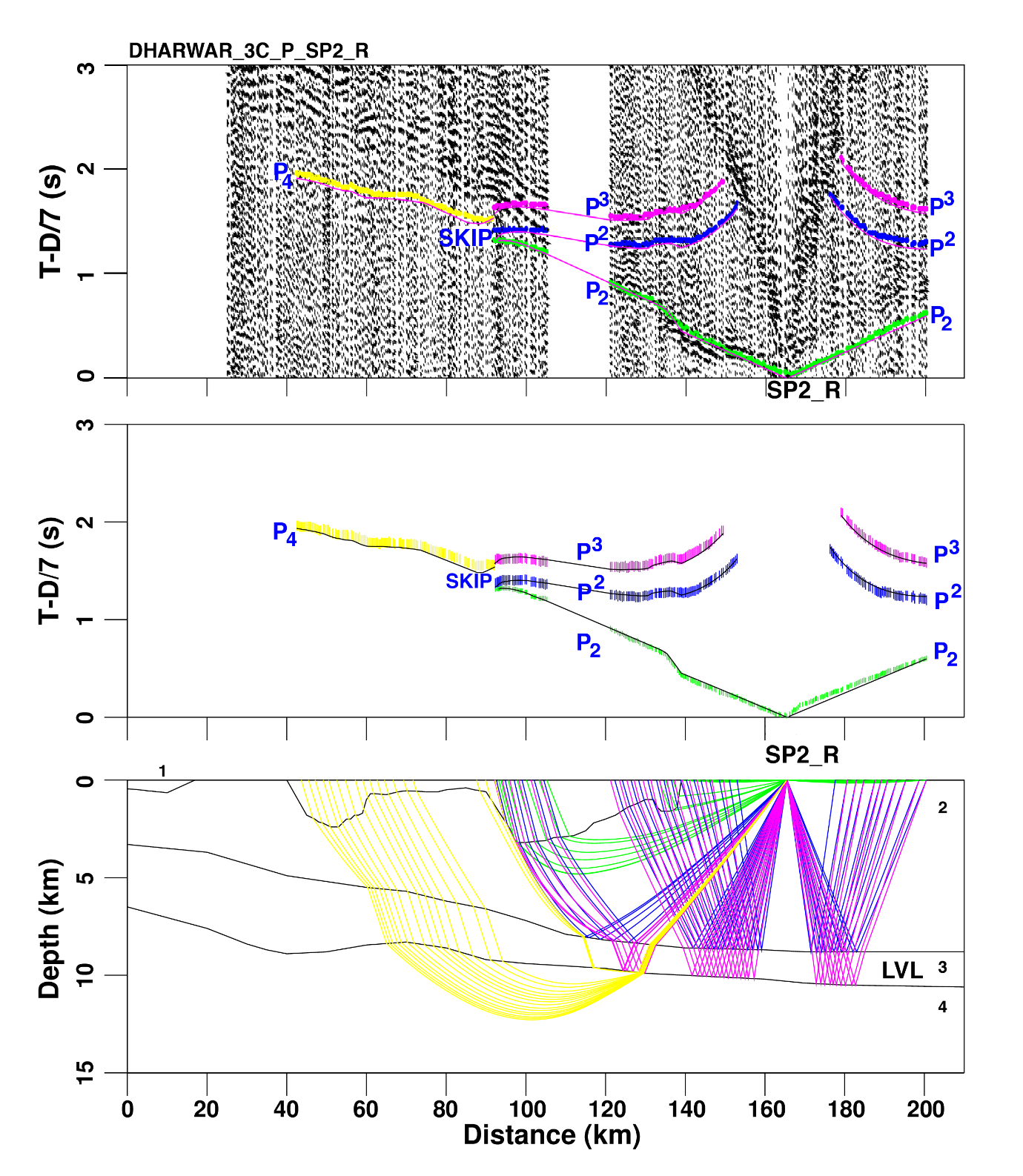}
\vspace{0.8em}
\textbf{(b) Deep SP1 observed gather and interpreted picks}\par
\includegraphics[width=0.96\textwidth,trim=0 790bp 0 0,clip]
  {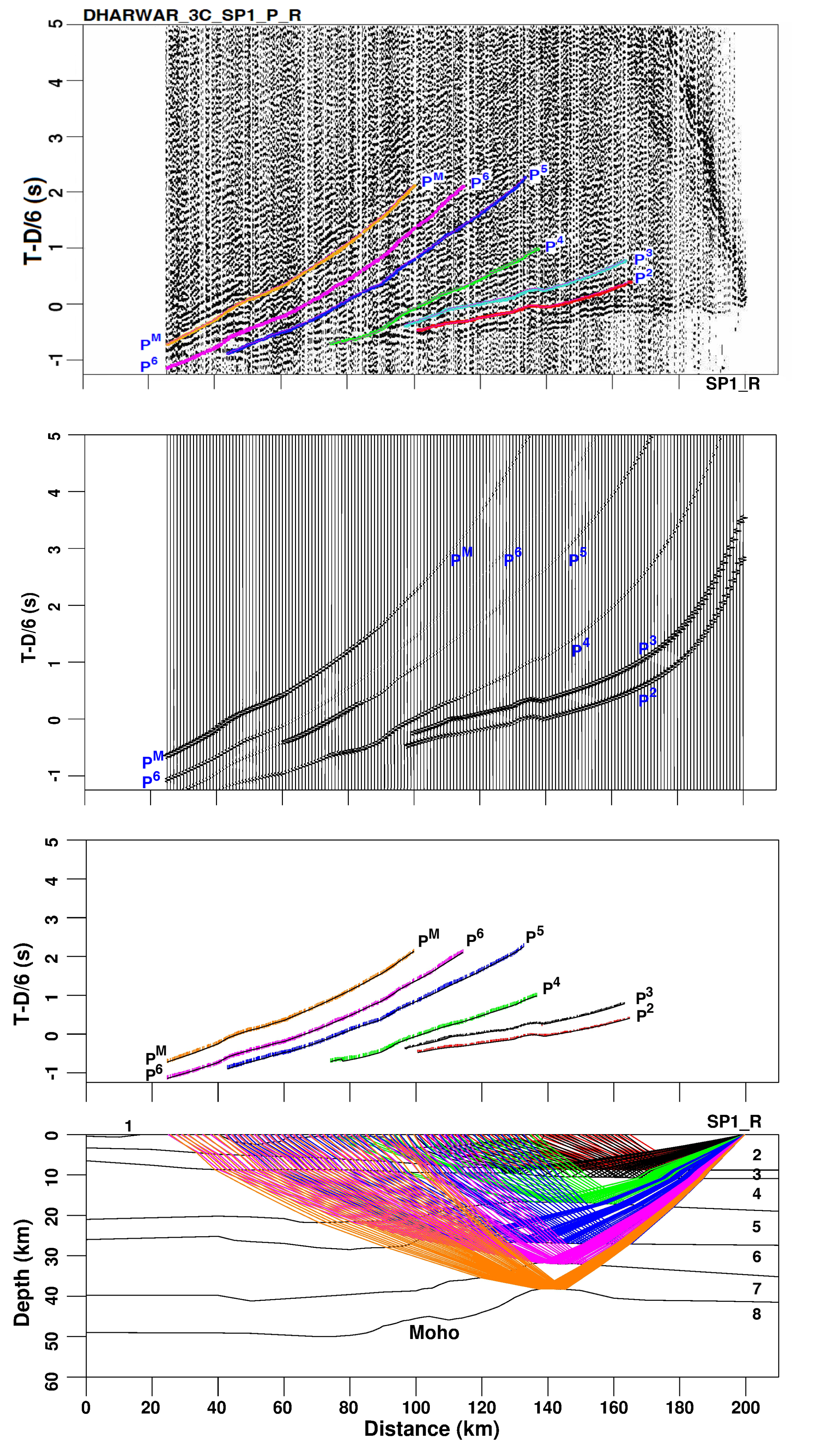}
\caption{Representative picked observations inherited from the thesis record
sections of \citet{kumar2022thesis} and \citet{beherakumar2022}. (a) The SP2
shallow section displays the $P_2$ turning phase, the ray-2.2 $P^{2}$ LVL-top
reflection, the ray-3.2 $P^{3}$ LVL-base reflection and the later-turning
$P_4$ branch. (b) The SP1 full-crust section displays the reflected
$P^{2}$--$P^{6}$ sequence together with the legacy $P^{M}$ label, which is PmP
in the present notation. Vertical coordinates are reduced time, $T-D/v_r$,
where $D$ is source--receiver distance and $v_r=7$ and 6~km\,s$^{-1}$ in (a)
and (b). The inherited horizontal ticks span 0--200~km profile distance at
20-km spacing, their numbered common axes being retained in Supplementary
Figs~S1 and S2. Coloured overlays are the inherited pick and branch
interpretation. These panels supply observational context and are not outputs
of the revised inversion.}
\label{fig:records}
\end{figure}

\begin{figure}[H]
\centering
\includegraphics[width=0.98\textwidth,height=0.84\textheight,keepaspectratio]{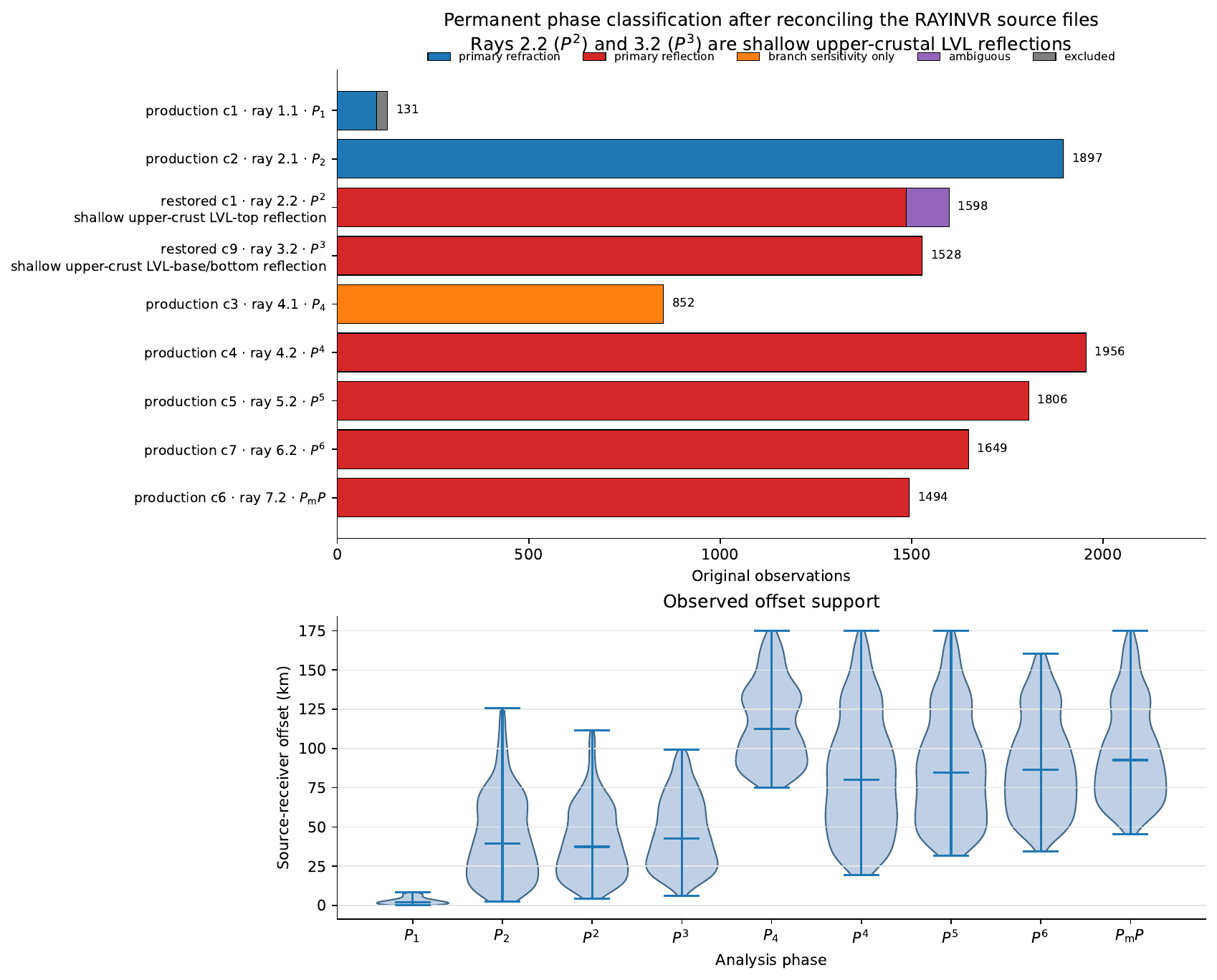}
\caption{Permanent phase classification after reconciling the two
\textsc{rayinvr} source-file generations. The upper panel gives record counts
with their analysis disposition and the lower panel absolute-offset support.
Ray 2.2 is $P^{2}$, reflecting from the bottom of layer 2 and hence the LVL
top; ray 3.2 is $P^{3}$, reflecting from the bottom of layer 3 and hence the
LVL base. The 1,598 ray-2.2 records comprise 1,486 primary $P^{2}$ picks plus
112 ambiguous $P^{2}/P_2$ records excluded from the primary set. Ray 4.1 is the
distinct later-turning $P_4$ branch of 852 observations, used only in the
labelled sensitivity analysis. Primary-refraction and primary-reflection
identifiers do not overlap.}
\label{fig:phaseaudit}
\end{figure}

Record sections inherited from SP2 (upper crust) and SP1 (full crust) are
retained purely as observational context; the deep-crustal interpretation was
presented by \citet{beherakumar2022}, the upper-crustal interpretation by
\citet{kumarbehera2023}, and both derive from \citet{kumar2022thesis}.
Figure~\ref{fig:records} reproduces legible crops with their pick overlays,
the complete legacy response and ray-diagram panels appearing as
Supplementary Figs~S1 and S2; none of this is output of the present inversion.
The analysis classification we adopt permanently is given in
Fig.~\ref{fig:phaseaudit}. Subscripts mark a direct or refracted phase together
with the layer in which it turns, superscripts a reflection from that layer's
base. Thus $P_1$ is the direct arrival, $P_2$ and $P_4$ turn in layers 2 and 4,
and $P^{2}$, $P^{3}$, $P^{4}$, $P^{5}$, $P^{6}$ and PmP reflect from the bases
of layers 2 through 7. Critically, \textsc{rayinvr} ray 2.2 is $P^{2}$, the
shallow upper-crustal reflection off the bottom of layer 2 and hence the LVL
top, while ray 3.2 is $P^{3}$, returning from the bottom of layer 3 and hence
the LVL base. The 852 records carried under code 3 in the later deep-crustal
file map to ray 4.1 and are therefore later-turning $P_4$ arrivals, not $P^{3}$
reflections at all. Table~\ref{tab:phases} collects the audited scheme.

\subsection{Disaggregating the merged phase codes}

What looked like code ambiguity arose simply because the two source-file
generations number their integer codes differently; the \textsc{rayinvr} ray
definitions themselves are unambiguous. The ray-control files supply the
permanent reading, in which ``L.1'' turns in layer L and ``L.2'' reflects from
that layer's base. We therefore keep both the original code and the source-file
generation in the audit ledger while using the ray number for classification.
Code 3 in the later file is ray 4.1, that is $P_4$, whereas the separately
restored LVL-base reflection is ray 3.2, that is $P^3$. Only the 112
observationally indistinguishable $P^2/P_2$ pairs remain genuinely ambiguous,
and they are held out of the primary inversion, entering solely in a labelled
sensitivity experiment.

\begin{table}[H]
\centering
\caption{Audited phase families. A \emph{subscript} marks a direct or refracted
phase and names the layer in which the ray turns; a \emph{superscript} marks a
reflection and names the layer whose base reflects. ``Current'' is the later
deep-crustal source file, ``restored'' the earlier files from which rays 2.2
and 3.2 were recovered. Counts and offsets describe primary-analysis
observations, except for $P_4$, which is held back for a labelled sensitivity
analysis.}
\label{tab:phases}
\small
\setlength{\tabcolsep}{3.5pt}
\begin{tabular}{@{}lllcrr@{}}
\toprule
Phase & Interpretation & Source/code & Ray & $n$ & Offset (km) \\
\midrule
$P_1$   & direct, layer 1                         & current/1  & 1.1 & 103  & 1.0--8.6 \\
$P_2$   & turning in layer 2                      & current/2  & 2.1 & 1897 & 2.4--125.6 \\
$P_4$   & later branch turning in layer 4         & current/3  & 4.1 & 852  & 75.1--175.1 \\
$P^{2}$ & LVL-top reflection (base of layer 2)    & restored/1 & 2.2 & 1486 & 4.4--108.1 \\
$P^{3}$ & LVL-base reflection (base of layer 3)   & restored/9 & 3.2 & 1528 & 6.2--99.2 \\
$P^{4}$ & reflection from base of layer 4         & current/4  & 4.2 & 1956 & 19.4--175.1 \\
$P^{5}$ & reflection from base of layer 5         & current/5  & 5.2 & 1806 & 31.5--175.1 \\
$P^{6}$ & reflection from base of layer 6         & current/7  & 6.2 & 1649 & 34.5--160.2 \\
PmP     & Moho reflection (base of layer 7)       & current/6  & 7.2 & 1494 & 45.4--175.1 \\
\bottomrule
\end{tabular}
\end{table}

\subsection{Picking uncertainty}
\label{sec:sigma}

The source files attach 25~ms to $P_1$, $P_2$ and $P^{2}$, and 50~ms to
$P^{3}$, $P^{4}$, $P^{5}$, $P^{6}$, PmP and the conditional $P_4$ branch. We
keep these audited figures as observation weights rather than building a fresh
error model out of the same withheld residuals that later serve for
evaluation. Whether they are adequate is settled directly in the nested outer
predictions of Section~\ref{sec:calib}, through standardized residuals and
empirical coverage. Any predictive scale inferred from those residuals is
treated as a post-validation diagnostic and is never fed back into the folds
that generated it.

\section{Method}

\subsection{Forward problem}

First arrivals follow from the eikonal equation
\begin{equation}
|\nabla T(\mathbf{x})| = \frac{1}{v(\mathbf{x})},
\label{eq:eikonal}
\end{equation}
solved on a grid of spacing $h$ by the fast-marching method of
\citet{sethian1996}. Since sources and receivers all lie on the free surface, a
single sweep per shot delivers the traveltime to every receiver of that shot.
We initialise the point source on a small circle of radius $r_0$ instead of one
cell and restore the source-to-circle time analytically as $r_0/v$, which
removes the dominant error a single-cell seed would introduce.

Rather than assume numerical accuracy we verify it against an analytic
solution. Where the vertical gradient is constant, $v(z)=v_0+kz$, the
surface-to-surface first arrival across offset $X$ is
\begin{equation}
T(X) = \frac{2}{k}\,\operatorname{arcsinh}\!\left(\frac{kX}{2v_0}\right).
\label{eq:analytic}
\end{equation}
Taking $v_0=5.3$~km\,s$^{-1}$ and $k=0.06$~s$^{-1}$, both representative of
this crust, the solver reproduces Eq.~\ref{eq:analytic} to 16.521~ms RMS at
$h=1$~km, 9.043~ms at $h=0.5$~km and 4.771~ms at $h=0.25$~km, the error falling
monotonically under refinement at an observed rate close to first order. We
adopt $h=0.25$~km in production, where numerical error accounts for roughly
four per cent of the variance budget set against a 25-ms picking error. One
caveat should be recorded: this benchmark is laterally homogeneous, so it
bounds the smooth-medium error but does not by itself bound the error in a
medium containing a low-velocity layer.

Computed naively, reflection traveltimes are the costly part of a wide-angle
forward problem, because every source--receiver pair demands its own solution.
Reciprocity permits a reorganisation, since sources and receivers all sit on
the surface. We solve Eq.~\ref{eq:eikonal} once from each point of a subsurface
grid and keep only the surface row of each solution. The resulting table gives
the traveltime between any surface point and any subsurface point, so that the
two-leg reflection time from a reflector $Z(x)$ becomes
\begin{equation}
T_{\mathrm{refl}}(S,R) = \min_{p}\left[\,T_p(S) + T_p(R)\,\right],
\qquad p = (x_p, Z(x_p)),
\label{eq:reflect}
\end{equation}
the minimisation being Fermat's principle in discrete form. A single table of
112 depth levels by 85 lateral columns then serves all seven shots, all 9,919
primary reflection picks and any candidate reflector geometry, and it is this
that makes the repeated refitting demanded by cross-validation affordable.

Numerical error in the reflection operator is likewise tested component by
component against finer or analytic references. At production spacings the
fast-marching field, lateral table interpolation, depth interpolation and
discrete Fermat search contribute RMS errors of 8.207, 2.929, 0.570 and
2.118~ms. Their descriptive root-sum-square is 8.986~ms, while the more
cautious sum of the four maximum absolute errors reaches 23.057~ms, equivalent
to 69~m of vertical two-way depth at 6~km\,s$^{-1}$ and thus below even the
smallest assigned pick uncertainty of 25~ms. A horizontal reflector placed at a
known 22.35~km is recovered at 22.34875~km, and the production central
0.01-km Jacobian departs from a fine reference by 0.0685 per cent RMS
(Supplementary Fig.~S9).

\subsection{The fixed production velocity field}

\begin{figure}[H]
\centering
\includegraphics[width=0.98\textwidth,height=0.86\textheight,keepaspectratio]{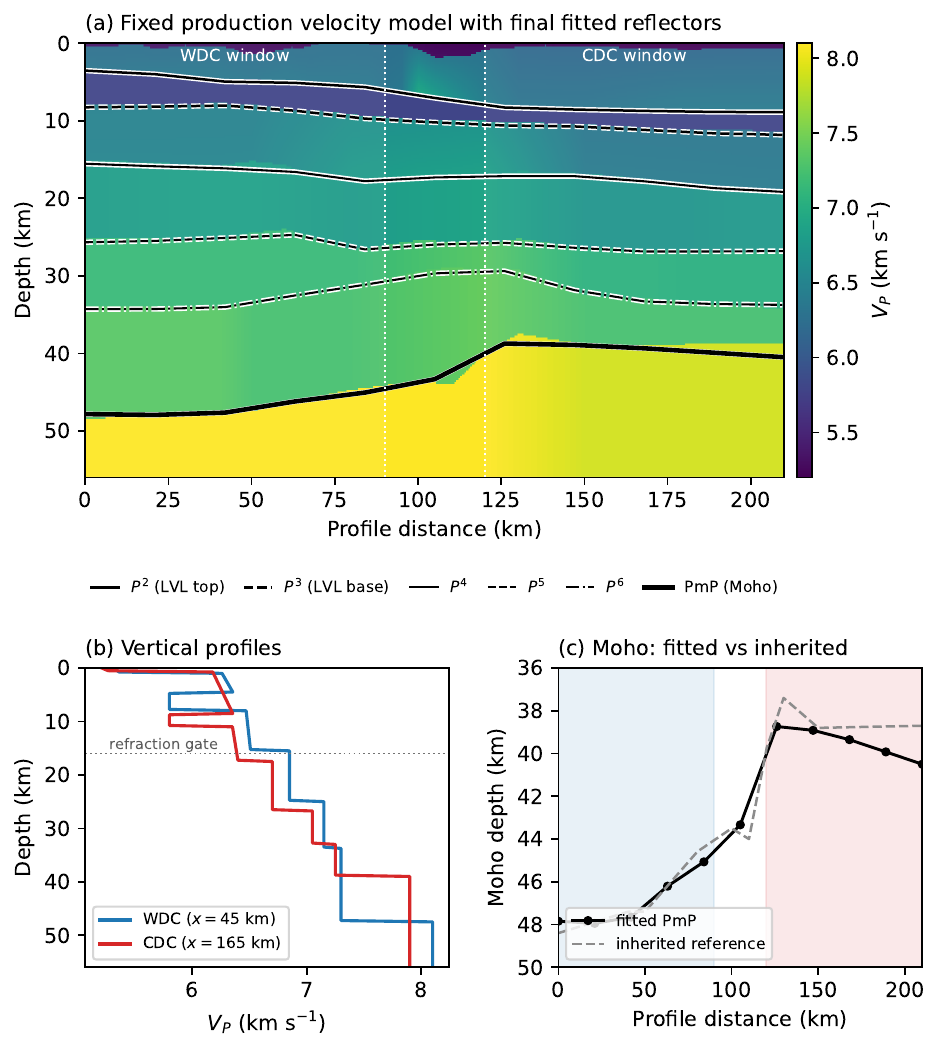}
\caption{The fixed production velocity field on which every reflector result is
conditional, derived from the ray-trace inversion of \citet{beherakumar2022}.
(a) $V_P$ with the final fitted six-reflection geometry overlaid and the WDC
$[0,90]$~km and CDC $[120,210]$~km averaging windows shaded. (b) Vertical
profiles through the WDC and CDC windows; the dotted line marks 16~km, the
greatest depth at which the present first arrivals clear the 0.02 coverage
gate, so that all velocity below it is inherited rather than measured here.
(c) The fitted PmP reflector against the inherited reference Moho. Velocities
nowhere fall below 5.2~km\,s$^{-1}$, so no weathered near-surface material is
represented, which bears directly on the systematic $P_1$ bias reported in
Section~\ref{sec:calib}.}
\label{fig:prodvel}
\end{figure}

Everything we report for the reflectors is conditional on one velocity field,
so we display it explicitly in Fig.~\ref{fig:prodvel}. It derives from the
ray-trace inversion of \citet{beherakumar2022}, itself following the approach
of \citet{zelt1988} and \citet{zelt1992}, and it is held fixed throughout
reflector fitting, validation and bootstrap resampling. Two features matter for
what follows. Velocities nowhere fall below 5.2~km\,s$^{-1}$, so neither this
field nor the refraction parameterisation described below contains any
weathered or sedimentary near-surface material. And the field is essentially
unconstrained by the present first arrivals below about 16~km depth, whereas
the Moho lies between 39 and 47~km; the depth conversion at Moho level
therefore inherits its velocity from the earlier study rather than measuring it
here.

\subsection{Inverse problem}

Refracted arrivals are inverted for a smooth velocity field using the 2,000
primary $P_1$ and $P_2$ observations alone. The 852 code-3 observations are ray
4.1, that is $P_4$, a later branch crossing the LVL and turning beneath it;
they are never pooled with first arrivals and are treated separately for
branch-specific LVL sensitivity. Slowness rides on nodes spaced 5~km laterally
and 1~km vertically and is interpolated bilinearly onto the forward grid, so
that no layers or interfaces are imposed at all. Rays are reconstructed after the event by steepest descent on the traveltime field, every receiver belonging to a shot being advanced in step. Since $T=\int u\,\mathrm{d}s$ and $u$ is bilinear in the node values, node $n$ contributes a Jacobian entry equal to the sum, taken over ray segments, of segment length weighted by the bilinear coefficient of node $n$ evaluated at the segment midpoint, which matches the parameterisation exactly.

There are 1,161 unknown node slownesses, bounded by velocities of
4.8--7.2~km\,s$^{-1}$. Within each training split a laterally uniform
reference $v(z)=v_0+kz$ is chosen from that split alone. At fixed
regularisation strength $\lambda$ the inversion minimises
\begin{equation}
\Phi_u(\mathbf{u}) =
\frac{1}{2}\left\|\mathbf{C}_d^{-1/2}
 [\mathbf{t}_{\rm obs}-\mathbf{F}(\mathbf{u})]\right\|^2+
\frac{1}{2}\left\|\mathbf{R}_s
 (\mathbf{u}-\mathbf{u}_{\rm ref})\right\|^2+
\frac{1}{2}\left\|\mathbf{R}_d
 (\mathbf{u}-\mathbf{u}_{\rm ref})\right\|^2 ,
\label{eq:refraction_objective}
\end{equation}
in which $\mathbf{C}_d$ is diagonal and holds the original per-pick variances,
$\mathbf{R}_s$ contains first spatial differences, and $\mathbf{R}_d$ applies
coverage-weighted damping towards the training-only reference. Regularisation
scales and coverage weights are fixed once from the weighted Jacobian at that
reference; they are neither recomputed after updates nor divided out afresh for
each candidate. Every backtracking trial evaluates
Eq.~\ref{eq:refraction_objective} down to a step of $1/128$. A fit terminates
when the relative penalised-objective decrease falls to $10^{-4}$ or below,
when the maximum slowness update falls to $10^{-5}$~s\,km$^{-1}$ or below, or
when all eight declared full-objective line-search steps raise the objective.
A stop forced by an iteration ceiling would invalidate the fit for validation
purposes, and none occurred.

The layered model against which our result is compared came from ray-trace
inversion in the manner of \citet{zelt1988} and \citet{zelt1992} and is
reported by \citet{beherakumar2022}. Reflector depths here ride on eleven nodes
at 21-km spacing, and because all six interfaces are fitted at once there are
66 free depth parameters. Writing $\mathbf{z}$ for the stacked depth nodes and
$\mathbf{z}_{0}$ for flat interfaces at their reference mean depths, the
complete objective is
\begin{align}
\Phi_z(\mathbf{z}) ={}&
\operatorname{mean}_i\!\left[
 \left(\frac{t_i^{\mathrm{obs}}-t_i^{\mathrm{refl}}(\mathbf{z})}
 {\sigma_i}\right)^2\right]
+\lambda_s^2\operatorname{mean}\!\left[
 \left(\frac{\mathbf{D}_1\mathbf{z}}{s_s}\right)^2\right]
\nonumber\\
&+\lambda_c^2\operatorname{mean}\!\left[
 \left(\frac{\mathbf{D}_2\mathbf{z}}{s_c}\right)^2\right]
+\lambda_r^2\operatorname{mean}\!\left[
 \left(\frac{\mathbf{z}-\mathbf{z}_0}{s_r}\right)^2\right],
\label{eq:reflector_objective}
\end{align}
where $\mathbf{D}_1$ and $\mathbf{D}_2$ are within-interface first- and
second-difference operators and $s_s=s_c=s_r=1$~km. Base coefficients
$(\lambda_s,\lambda_c,\lambda_r)=(0.10,0.15,0.05)$ are scaled by the
nested-validation candidate multiplier, so that the final full-data multiplier
of 3 yields effective coefficients $(0.30,0.45,0.15)$. Normalising each block
by its own row count keeps the coefficients independent of how many picks or
nodes there are, in preference to dividing by a model-dependent norm that can
quietly cancel the intended weight.

Gauss--Newton updates employ a central 0.01-km finite-difference traveltime
Jacobian and an untruncated least-squares solve, in the spirit of the
formulation given by \citet{tarantola2005}. Each candidate is projected so as
to keep adjacent interfaces at least 0.25~km apart, and a backtracking line
search accepts it only when the complete objective of
Eq.~\ref{eq:reflector_objective} falls. Iteration halts once the relative
objective change drops below $10^{-4}$ and the maximum node update below
0.01~km simultaneously, or once every declared line-search step has been
tried, with at most 12 accepted iterations permitted.

Damping in the refraction inversion is weighted by the ray coverage of each
node, read from the diagonal of the weighted Jacobian, and pulls towards the
one-dimensional starting model rather than towards zero. We mask reported
velocities wherever normalized reference-model coverage falls below 0.02;
values beyond that gate remain regularised extrapolations and are not
interpreted.

\subsection{Phase-specific sensitivity of the low-velocity layer}

Our LVL analysis retains the lateral anomalies of the independently selected
six-reflector geometry at regularisation multiplier 3. Candidate models impose
a uniform shift on the $P^{2}$ interface and alter the mean-thickness parameter
governing the $P^{2}$--$P^{3}$ separation, while preserving the fitted lateral
anomaly of that separation. It is worth being explicit that this scan parameter
is referenced to the nodal mean thickness of the final geometry, 3.224~km, and
is therefore not numerically interchangeable with the exact profile-window
integral of 3.163~km quoted later. We scan top offsets from $-1.0$ to
$+1.0$~km in 0.25-km steps, mean-thickness parameters from 1.75 to 5.50~km in
0.25-km steps, and core velocities from 5.2 to 6.6~km\,s$^{-1}$ in
0.1-km\,s$^{-1}$ steps. Three low-velocity profiles are tried: a sharp constant
layer, a linear internal recovery of 0.30~km\,s$^{-1}$, and a cosine-smoothed
recovery of equal magnitude. Continuous interpolation between the velocities
above and below the interval constitutes the no-internal-low-velocity model.
Altogether the grid holds 6,624 candidates.

For $P^{2}$ and $P^{3}$ we retain the exact two-dimensional
fast-marching/Fermat prediction of the final model as an anchor, representing a
candidate perturbation only through the difference between finite-offset
horizontally layered ray calculations for that candidate and for the reference.
Constant and linearly varying velocity segments are integrated analytically and
the ray parameter is found by bisection, following the layered formulation of
\citet{zelt1988}. The reference prediction is thus fully two-dimensional
whereas the candidate differential is a local one-dimensional approximation.
For a sharp constant LVL the zero-offset contribution is
$\tau_{\rm LVL}=2h/v$, so every pair $(h,v)$ sharing the same $h/v$ is exactly
equivalent for that datum, and only finite-offset moveout can narrow the
trade-off.

The 852 code-3 observations, being \textsc{rayinvr} ray 4.1, are never passed
to the first-arrival objective. We instead build a ray-parameter curve for every complete-shot propagation corridor, oblige its turning point to lie somewhere between the LVL base and the floor of layer 4, and solve for one nuisance static in each shot. What this operator quantifies is branch reachability together with moveout, conditional on a layered model averaged along the corridor; it neither traces branches in two dimensions nor preserves any absolute-time information.

Our preferred low-velocity candidate minimises the sum of shot-clustered
standardised phase misfits and the fixed final-geometry regularisation.
Independently of that, we label a candidate rule-based admissible only where
four requirements hold at once: its $P^{2}$ and $P^{3}$ RMS values sit at or
beneath the strict nested outer-phase figures of 35.861 and 47.962~ms, each of
the 852 $P_4$ offsets proves reachable, every $P_4$ ray turns within layer 4,
and its conditional RMS does not exceed a single 50-ms pick uncertainty. We note as a limitation that these rules gate only the two
LVL boundary phases and the $P_4$ branch, and place no constraint on the four
deeper reflections, whose predictions the LVL nevertheless perturbs; the
consequences are quantified in Section~\ref{sec:lvlresults}. The 112 ambiguous
$P^{2}/P_2$ records enter only a second calculation and are never
double-counted. Diagnostic $\Delta Q=2$, 5 and 10 contours are drawn for
orientation and are not mapped onto confidence levels.

\subsection{Calibration}
\label{sec:method_calib}

Validation is grouped by complete shot gather. For reflection validation the
velocity field and reciprocal table stay fixed throughout, so the protocol is
leakage-free with respect to reflector geometry, preprocessing and
regularisation selection, conditional on that forward model. In each of seven
outer folds one shot is set aside before any tuning at all. Regularisation is
then selected by leave-one-shot-out validation on the remaining six shots
alone, the model is refitted to those six from the same neutral start, and the
untouched outer shot is predicted exactly once. Pooling the outer predictions
therefore estimates out-of-sample performance without the test shot having
influenced regularisation selection, which is the specific failure that
\citet{varma2006} and \citet{cawley2010} identified in unnested procedures and
that \citet{roberts2017} showed to be aggravated by spatial structure. Smaller
weights ordinarily lower in-sample misfit and are not preferred on that ground,
and the candidate grid is extended geometrically whenever an inner optimum
lands on a boundary.

For the refraction inversion the final grid is
$\{0.25,0.5,1,2,4,8,16,32,64,128\}$. Within an outer fold each candidate is fitted to five complete shots and validated against the sixth, the cycle running over all six outer-training shots so that their standardized residuals may be pooled; the untouched outer shot contributes nothing whatever to this selection. Once the seven
nested outer predictions are complete, a separate seven-shot leave-one-shot-out
tuning pass selects the $\lambda$ used for the all-data model; those tuning
predictions serve only for that final selection and are never substituted for
the nested outer performance estimate.

Because only seven shot groups exist, we report the outer residual coverage as
empirical pick-level coverage conditional on this acquisition and model class,
and explicitly not as a distribution-free cluster-conformal guarantee of the
kind formalised by \citet{vovk2005}. Shot-to-shot spread is kept separate.
Dependence inside a gather is examined with a non-circular moving-block
bootstrap in receiver coordinate, combining the resampling principle of
\citet{efron1979} with the block construction of \citet{kunsch1989}. Blocks are drawn independently inside each shot, with replacement, and selecting a receiver carries with it every phase recorded there together with its exact bootstrap multiplicity. Final reflection residuals have an integral length scale of 9.416~km, and that is what motivates our predeclared 10-km primary block length. We use 500
primary replicates plus 200 replicates each at 5 and 20~km. Every replicate
contains all seven shots, recomputes its own flat reference from the resample,
and is fitted at the final multiplier of 3. Delete-one-shot estimates are
reported apart, as a seven-member sensitivity diagnostic.

For velocity--depth ambiguity we construct a full nonlinear perturbation ensemble in place of a single common scale factor. Six cosine-mask families, every one of them predeclared, act in turn upon the common deep crust, upon WDC alone, upon CDC alone, upon middle and lower crust along separate axes, upon the transition, and upon the near-Moho interval, each across $-3$ to $+3$ per cent; a true $3\times3$ factorial additionally combines
$-2$, 0 and $+2$ per cent middle- and lower-crust perturbations. This yields 58
ledger entries and 47 unique velocity fields. Every unique field prompts a complete rebuild of the fast-marching table followed by a fresh simultaneous fit of all six interfaces to the identical 9,919 reflections at multiplier 3. A model counts as admissible only
if its interfaces stay ordered, its fit converges, and neither its reflection
RMS nor its standardized MSE exceeds the corresponding untouched nested-outer
value of 67.77~ms and 2.068. Should a perturbation encroach upon the supported refraction mask, we additionally require the shift it induces in the predictions for those same 2,000 $P_1$--$P_2$ observations to remain beneath the nested-outer refraction RMS. Comparing an in-sample refit against a held-out threshold makes
this gate deliberately permissive, which is why the admissible range reported
below should be read as a conservative envelope rather than a tight bound. The
$\pm3$ per cent amplitude is itself a design choice, not a measured bound, and
we return to its consequences in Section~\ref{sec:claims}.

Province means are not simple averages over the eleven reflector nodes. Our
publication estimand is the exact length-weighted integral of the linearly
interpolated Moho across WDC $[0,90]$~km and CDC $[120,210]$~km, the contrast
being the former minus the latter, and the same operator is applied in the
bootstrap and synthetic experiments alike. Targeted recovery uses the actual
1,494-pick PmP geometry, the production start and multiplier 3. Fifteen imposed
Moho transitions pair centres at 100, 110 and 120~km with widths of 0, 20, 40,
60 and 100~km, each inverted noise-free and then in 50 independent pick-noise
realisations, so that bias, random spread and coverage rest on 15 noise-free
fits together with 750 noisy inversions, none of them retuned. Accepted
non-overlapping province-window tests displace the inner limits outward by 10
and 20~km and inward by 10~km; the predeclared 20-km inward case is rejected
because the province windows would then overlap. These experiments delimit
design-class and boundary sensitivity, and because they cover neither every
possible velocity field nor anisotropy nor three-dimensional structure, they
are never called total uncertainty.

\section{Results}

\subsection{Depth reached by the refracted arrivals}

Turning depths from the final traced rays extend to 20.405~km along the single
deepest paths, yet their 95th percentile is a mere 6.04~km. Normalized coverage
clearing the 0.02 interpretation gate embraces 95.3 per cent of model nodes at
the surface, 48.8 per cent at 8~km and 23.3 per cent at 14~km, with just two
nodes passing at 16~km and none at 17~km (Fig.~\ref{fig:refr}). Refracted
arrivals accordingly constrain the upper crust and cannot possibly reach a Moho
between 38 and 48~km. This does not leave the upper crust resting on
refractions alone, because $P^{2}$ and $P^{3}$ are reflections from the LVL top
and base; in the selected phase-specific model their mean depths are 6.684 and
9.684~km, though both vary laterally. Conversely every interface below the
reach of the first arrivals, the Moho included, is constrained by reflections.
The two data families are complementary, not interchangeable.

The contextual record sections of Supplementary Figs~S1 and S2 display the
phase gap between $P_2$ and $P_4$, which is consistent with there being no
$P_3$ turning branch; the quantitative LVL tests below nevertheless rest on
matched $P^{2}/P^{3}$ reflections and a separately declared conditional $P_4$
operator rather than on extrapolation across that gap. Deep refraction coverage
is strikingly asymmetric. SP1 at the north-eastern end records $P_2$ out to
125.63~km and holds the deepest traced ray at 20.405~km, whereas SP7 at the
south-western end records $P_2$ only to 78.68~km and its rays reach 6.07~km.
Refraction constraint at depth is thus concentrated near the profile centre,
and every model value beyond the coverage gate is treated as regularised
extrapolation.

\subsection{Phase identification}
\label{sec:phaseid}

Each of the six reflected phases was tested against every final fitted boundary
at the independently selected multiplier 3 (Supplementary Fig.~S5). In every
row the correct boundary is the unique RMS minimum. Diagonal RMS values for
$P^{2}$, $P^{3}$, $P^{4}$, $P^{5}$, $P^{6}$ and PmP are 23.448, 43.140,
60.113, 63.957, 43.464 and 57.678~ms, and the excess over the diagonal for the
nearest wrong boundary is 581.334, 504.410, 629.699, 841.959, 773.024 and
1671.491~ms respectively; PmP, for instance, fits the Moho at 57.678~ms against
1729.169~ms for the nearest alternative.

The logical standing of this test needs stating plainly, because it is weaker
than the numbers suggest. Each boundary was fitted using the very phase then
shown to prefer it, so the diagonal cannot fail to be the minimum, and the
large excesses chiefly record that the six interfaces are widely separated in
traveltime. What the test therefore establishes is detectability rather than
identification: interface separations exceed the pick uncertainties by so much
that a gross mis-assignment would have been conspicuous. It is an in-sample
diagnostic and not a held-out calibration, and the later-turning $P_4$ branch
forms no part of the six-reflection matrix.

Refracted phases possess no reflector and are instead fitted as first arrivals.
In the final all-data solution their in-sample RMS residuals are 37.41~ms
across the 103 $P_1$ picks and 14.61~ms across the 1,897 $P_2$ picks, pooling
to 16.57~ms, with corresponding biases of $+34.70$, $+1.45$ and $+3.16$~ms.
None of these in-sample values is used as predictive calibration.

\subsection{Calibration of the picking uncertainty}
\label{sec:calib}

\begin{figure}[H]
\centering
\includegraphics[width=0.98\textwidth,height=0.84\textheight,keepaspectratio]{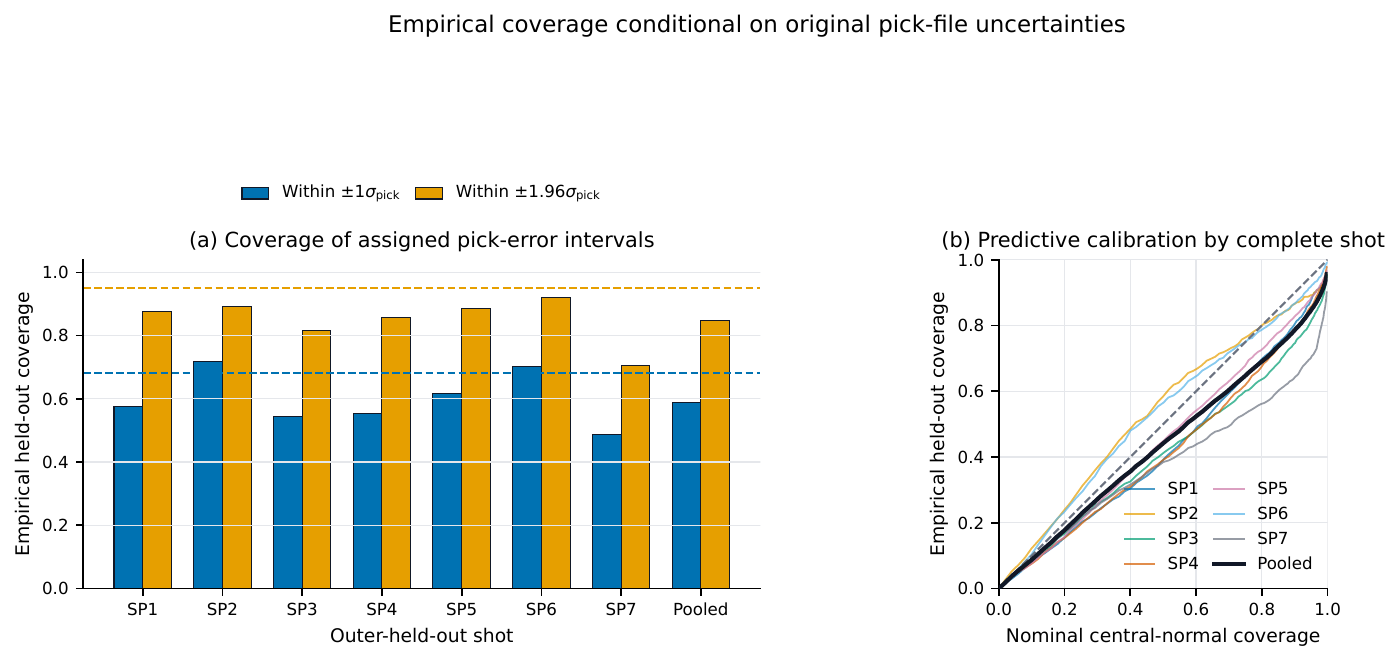}
\caption{Leakage-free nested leave-one-complete-shot-out prediction for all
9,919 primary reflections, conditional on the fixed velocity field and
reciprocal table. (a) Empirical coverage within $\pm1\sigma_{\mathrm{pick}}$
and $\pm1.96\sigma_{\mathrm{pick}}$ for each untouched shot and for the pooled
predictions, dashed lines marking the nominal 0.68 and 0.95 levels.
(b) Shot-wise and pooled predictive-calibration curves comparing empirical
central coverage with nominal central-normal coverage, the dashed diagonal
being ideal calibration. Pooled RMS is 67.77~ms, bias 3.28~ms, standardized MSE
2.068, and empirical coverage 0.590 and 0.848 at the nominal 68 and 95 per cent
limits. Every primary reflection receives exactly one untouched outer
prediction. Table~\ref{tab:perphase} resolves these pooled figures by phase.}
\label{fig:calib}
\end{figure}

Nested complete-shot validation yields exactly one prediction for each of the
9,919 primary reflections, with no overlap whatever between an outer test shot
and any inner training or selection set. Pooled held-out RMS is 67.77~ms and
the bias 3.28~ms. Judged against the pick-file uncertainties, the standardized mean-squared residual comes out at 2.068 in place of unity, while the empirical fractions falling inside nominal 68 and 95 per cent Gaussian limits are 0.590 and 0.848 (Table~\ref{tab:validation}; Fig.~\ref{fig:calib}). The assigned
uncertainties consequently understate predictive dispersion. Coverage varies
materially between shots as well: the worst withheld gather returns 97.35~ms
RMS with only 0.706 coverage inside the nominal 95 per cent limit. These are
descriptive pick-level coverages over seven shot clusters, not formal
guarantees. Two of the seven outer refits, those withholding SP2 and SP7,
stopped after four and seven accepted iterations without meeting a declared
convergence condition; neither hit an iteration ceiling, but since SP7 is also
the worst-predicted gather we flag this explicitly rather than describe all
seven folds as converged.

\begin{table}[H]
\centering
\caption{Leakage-free nested complete-shot validation. ``All-data'' is
in-sample RMS for the separately selected final fit; every other quantity comes
from untouched outer-shot predictions alone. Coverage is the empirical pick
fraction satisfying
$|t_{\mathrm{obs}}-t_{\mathrm{pred}}|/\sigma_{\mathrm{pick}}\leq 1$ or 1.96.}
\label{tab:validation}
\small
\setlength{\tabcolsep}{3.5pt}
\begin{tabular}{@{}lrrrrrrr@{}}
\toprule
Data & $n$ & All-data & Held-out & Bias & Std.\ MSE & Cov.\ 68 & Cov.\ 95 \\
 & & \multicolumn{3}{c}{RMS or bias (ms)} & & & \\
\midrule
Reflections & 9,919 & 51.40 & 67.77 & 3.28 & 2.068 & 0.590 & 0.848 \\
$P_1$--$P_2$ & 2,000 & 16.57 & 102.60 & 50.14 & 16.844 & 0.283 & 0.533 \\
\bottomrule
\end{tabular}
\end{table}

Pooling across phases conceals a great deal, so Table~\ref{tab:perphase}
resolves the same untouched predictions phase by phase. The picture is markedly
heterogeneous. For $P^{3}$ and $P^{6}$ the assigned 50~ms is close to adequate,
their standardized MSE values being 0.920 and 1.124, whereas $P^{4}$ and
$P^{5}$ are understated by factors near three. Most importantly for this paper,
PmP, on which every Moho statement depends, predicts withheld shots at
72.756~ms with standardized MSE 2.117 and nominal-95 coverage 0.841. It is that
figure, rather than the pooled 67.77~ms, which properly scales Moho
uncertainty, and the blanket claim that assigned uncertainties are too small
holds on average while being false for two of the six phases.

\begin{table}[H]
\centering
\caption{Per-phase decomposition of the same untouched outer-shot predictions
summarised in Table~\ref{tab:validation}. Assigned $\sigma$ is 25~ms for
$P^{2}$ and 50~ms for the remaining phases. Standardized MSE would be unity
were the assigned uncertainties correct.}
\label{tab:perphase}
\small
\setlength{\tabcolsep}{5pt}
\begin{tabular}{@{}lrrrrrr@{}}
\toprule
Phase & $n$ & Held-out RMS (ms) & Bias (ms) & Std.\ MSE & Cov.\ 68 & Cov.\ 95 \\
\midrule
$P^{2}$ & 1,486 & 35.861 &  5.01 & 2.058 & 0.504 & 0.821 \\
$P^{3}$ & 1,528 & 47.962 & 10.00 & 0.920 & 0.773 & 0.955 \\
$P^{4}$ & 1,956 & 86.158 &  9.99 & 2.969 & 0.569 & 0.782 \\
$P^{5}$ & 1,806 & 85.049 & $-1.66$ & 2.893 & 0.494 & 0.781 \\
$P^{6}$ & 1,649 & 53.008 & $-5.67$ & 1.124 & 0.674 & 0.929 \\
PmP     & 1,494 & 72.756 &  1.75 & 2.117 & 0.538 & 0.841 \\
\midrule
Pooled  & 9,919 & 67.766 &  3.28 & 2.068 & 0.590 & 0.848 \\
\bottomrule
\end{tabular}
\end{table}

The regularisation multiplier is searched over $\{0.1,0.3,1,3,9\}$ relative to
fixed slope, curvature and reference coefficients, and the grid brackets every
inner optimum selected. Four of the seven outer folds choose multiplier 3 and
three choose multiplier 1, while the separate full-data leave-one-shot-out
tuning selects multiplier 3, its standardized validation MSE being 1.913
against 2.008 at multiplier 1 and 3.262 at multiplier 9 (Supplementary
Figs~S6 and S7). The corresponding effective slope, curvature and reference
coefficients are 0.30, 0.45 and 0.15.

\begin{figure}[H]
\centering
\includegraphics[width=0.98\textwidth,height=0.84\textheight,keepaspectratio]{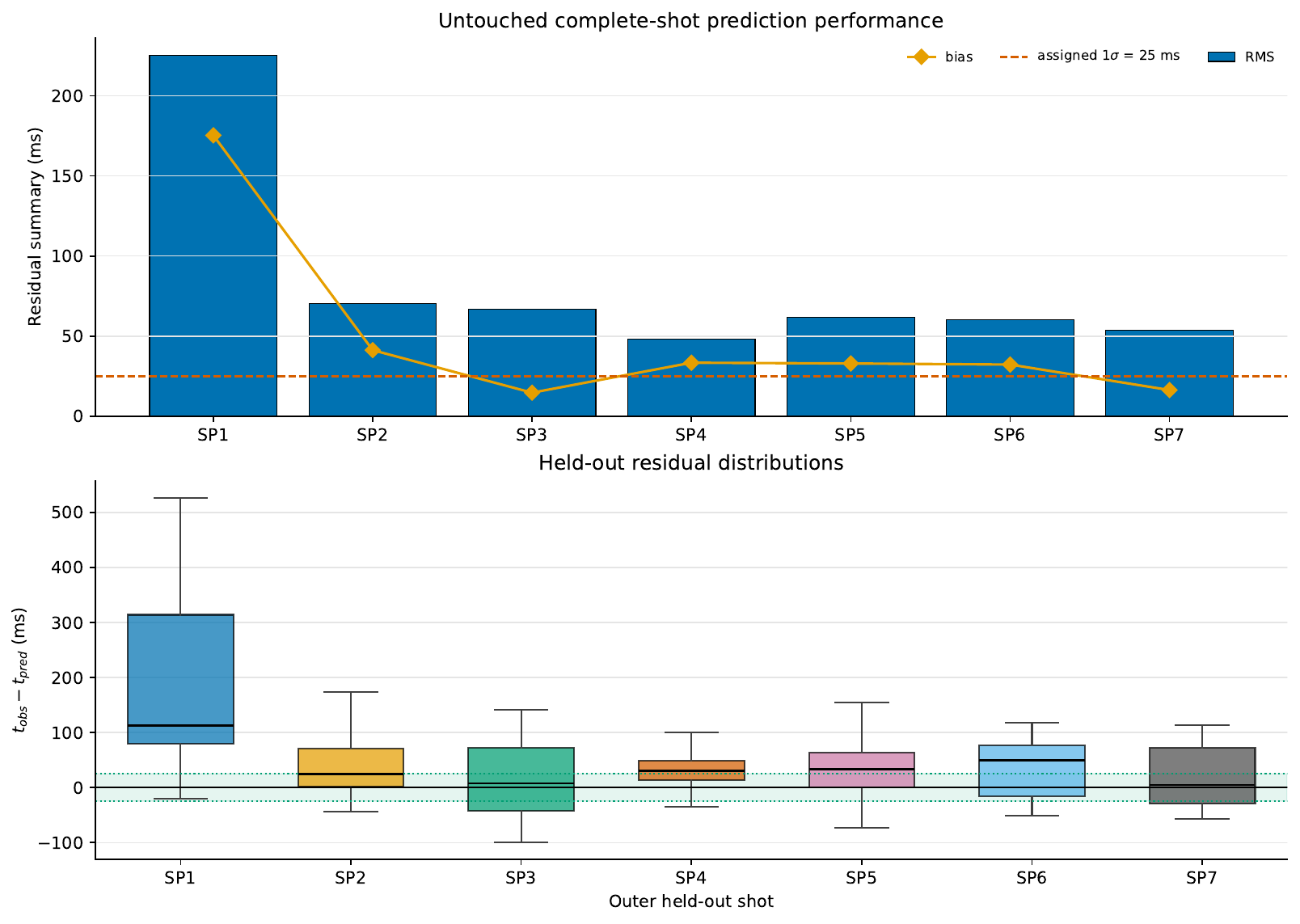}
\caption{Leakage-free nested leave-one-complete-shot-out prediction for the
2,000 primary $P_1$--$P_2$ observations. Bars and diamonds give RMS and bias
for each untouched shot and the dashed line the assigned 25-ms
one-standard-deviation uncertainty; box plots give held-out residual
distributions. Pooled untouched-shot RMS is 102.60~ms with $+50.14$~ms bias,
and the empirical fractions within one and 1.96 assigned standard deviations
are 0.283 and 0.5325. SP1 alone contributes 225.07~ms while the remaining six
shots span 48.04--70.25~ms, indicating that transfer to an unsampled aperture
dominates the pooled figure. The 16.57-ms all-data residual is an in-sample fit
and not predictive calibration.}
\label{fig:refrperf}
\end{figure}

For the 2,000 $P_1$--$P_2$ observations the seven nested inner selections are
$\lambda=4,4,8,1,1,1$ and 2 for outer SP1 through SP7, every optimum lying
interior to the searched grid (Supplementary Figs~S3 and S4). Untouched outer
predictions give 102.60~ms pooled RMS, $+50.14$~ms bias and standardized MSE
16.844, with empirical central coverage of only 0.283 at one assigned standard
deviation and 0.5325 at 1.96. A separate full-data tuning pass selects the
bracketed value $\lambda=1$, whose 96.01-ms tuning RMS is not substituted for
the nested estimate (Fig.~\ref{fig:refrperf}).

Two distinct effects lie behind this failure, and conflating them would
misstate what has been shown. The first is aperture extrapolation. SP1 alone
contributes 225.07~ms RMS while the other six shots range from 48.04 to
70.25~ms, and SP1 is precisely the shot recording the longest offsets and
holding the deepest rays, so withholding it removes the sole constraint on the
far aperture and then demands prediction there. With shots positioned at the
ends of a single line, seven gathers are not exchangeable, and the pooled
figure therefore measures transfer to an unsampled aperture at least as much as
it measures pick-error miscalibration. The second effect is a parameterisation
deficiency. Held out, the 103 $P_1$ direct arrivals carry a $+94.39$~ms bias
with coverage of exactly 0.000 at one assigned standard deviation and 0.097 at
1.96, and their in-sample bias is already $+34.70$~ms in the same direction.
Since the node velocities are bounded below at 4.8~km\,s$^{-1}$ and the
production field never falls below 5.2~km\,s$^{-1}$, no weathered near-surface
material can be represented at all, so the solver is forced to predict early
and the residuals are systematically one-sided. That, rather than dependence
among picks, is the most economical reading of the strongly positive pooled
bias of $+50.14$~ms. It follows that the original 25-ms pick errors do not
describe complete-shot transfer uncertainty for this parameterisation, but the
demonstration is not by itself proof that the picks were assigned too
optimistically.

\subsection{Starting-geometry sensitivity}

At the selected multiplier, a fit launched from six flat training-only
references and one launched from the existing \textsc{rayinvr} boundaries reach
effectively the same solution, their all-reflection RMS values being 51.396 and
51.422~ms. Across all 66 nodes the RMS and maximum absolute final depth
differences are 0.00150 and 0.00764~km, and the WDC--CDC Moho contrast shifts
by 0.000426~km (Supplementary Fig.~S8). This is a two-start sensitivity check
and nothing more; it is not evidence of independence from velocity,
regularisation, parameterisation or the broader model class. Our primary model
uses the flat start.

\subsection{Moho depth and the separate uncertainty axes}

\begin{figure}[H]
\centering
\includegraphics[width=0.98\textwidth,height=0.84\textheight,keepaspectratio]{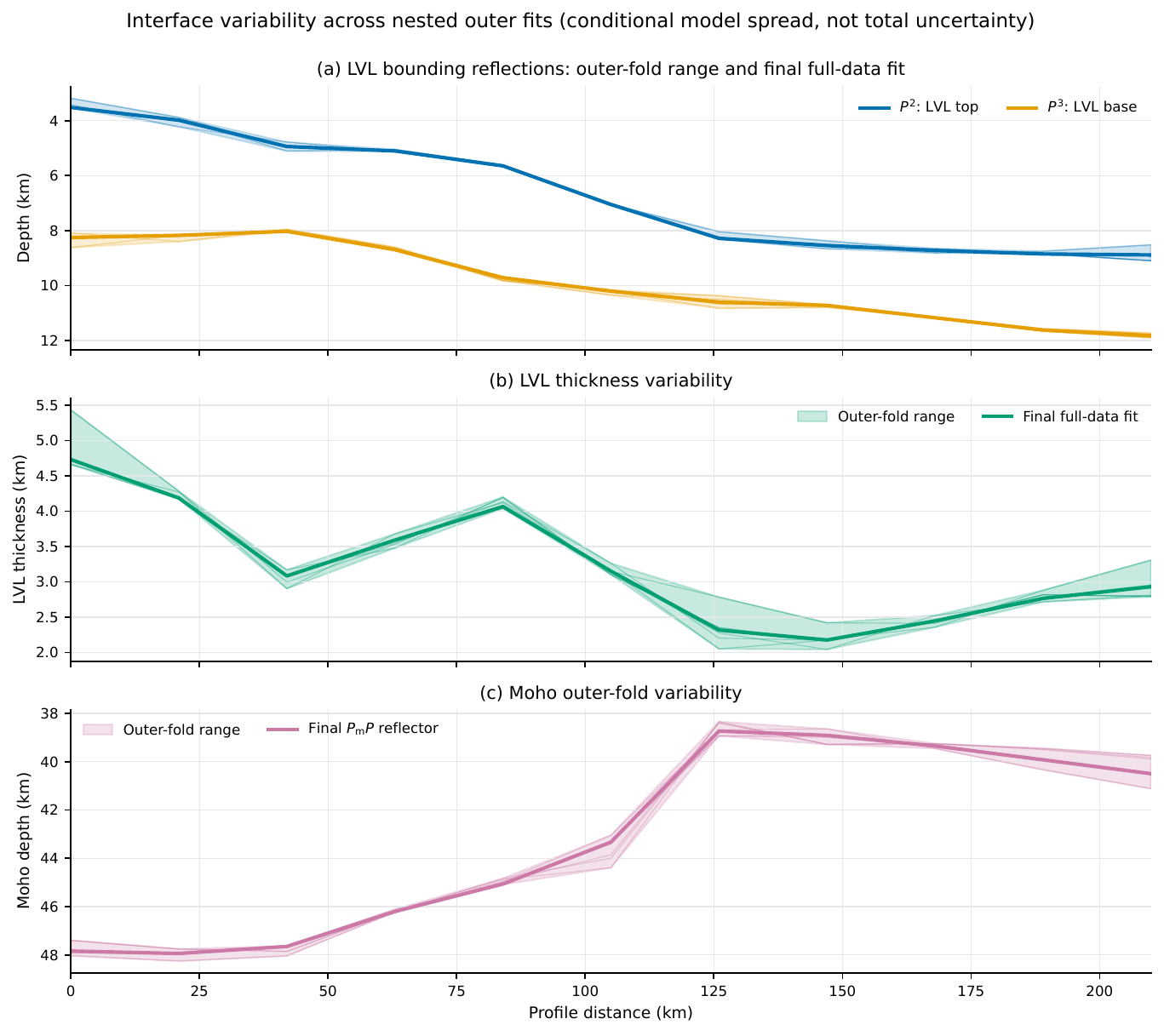}
\caption{Final fitted six-reflection interface geometry at the independently
selected multiplier 3. (a) The $P^{2}$ LVL-top and $P^{3}$ LVL-base reflectors
with the range across seven outer complete-shot fits. (b) LVL thickness.
(c) The PmP Moho reflector. Heavy curves are the final all-data fit and light
envelopes the outer-fold ranges. Exact length-weighted Moho means over WDC
$[0,90]$~km and CDC $[120,210]$~km are 46.915 and 39.448~km, a raw contrast of
7.466~km. Outer-fold spread is conditional model sensitivity, not total
uncertainty.}
\label{fig:model}
\end{figure}

With the production velocity model fixed, the final flat-start inversion returns
exact length-weighted Moho means of 46.915~km over WDC ($x=0$--90~km) and
39.448~km over CDC ($x=120$--210~km), a raw difference of 7.466~km
(Fig.~\ref{fig:model}). These are averages of the linearly interpolated
reflector rather than simple means over the eleven nodes; the corresponding
nodal means, 46.942 and 39.485~km, differ by tens of metres and are reported
only to make the distinction concrete. Table~\ref{tab:budget} then separates
five diagnostics conditioning on different assumptions.

\begin{table}[H]
\centering
\caption{Moho estimands and separate uncertainty diagnostics, in km. WDC and
CDC are exact profile-window means over 0--90 and 120--210~km. The receiver
bootstrap is conditional on the fixed velocity field and operator and
represents dependence \emph{within} shots only. Delete-one-shot quantities are
descriptive for seven clusters but represent dependence \emph{between} shots
and are the wider, less assumption-laden statement. Synthetic intervals cover
the tested transition-design class at fixed velocity, and the velocity row is
an admissible model range rather than a confidence interval. These rows must
not be combined in quadrature or described as total uncertainty.}
\label{tab:budget}
\small
\setlength{\tabcolsep}{5pt}
\begin{tabular}{@{}p{6.2cm}rrr@{}}
\toprule
Diagnostic & WDC & CDC & WDC--CDC \\
\midrule
Fixed-velocity point estimate
  & 46.915 & 39.448 & 7.466 \\
10-km receiver bootstrap (within shot), percentile 95\%
  & 46.80--46.97 & 39.34--39.70 & 7.18--7.57 \\
Delete one complete shot, range
  & 46.391--46.945 & 39.382--39.571 & 7.008--7.534 \\
Delete one complete shot, descriptive SE
  & 0.455 & 0.142 & 0.398 \\
Delete-one jackknife, bias-corrected $\pm$ normal 95\%
  & $47.332\pm0.893$ & $39.386\pm0.279$ & $7.946\pm0.780$ \\
Synthetic bias-adjusted $\pm$ design-envelope 95\% half-width
  & $47.067\pm0.446$ & $39.240\pm0.207$ & $7.827\pm0.454$ \\
Admissible nonlinear velocity ensemble
  & 45.763--47.500 & 38.204--39.878 & 6.457--8.639 \\
\bottomrule
\end{tabular}
\end{table}

\begin{figure}[H]
\centering
\includegraphics[width=0.98\textwidth,height=0.84\textheight,keepaspectratio]{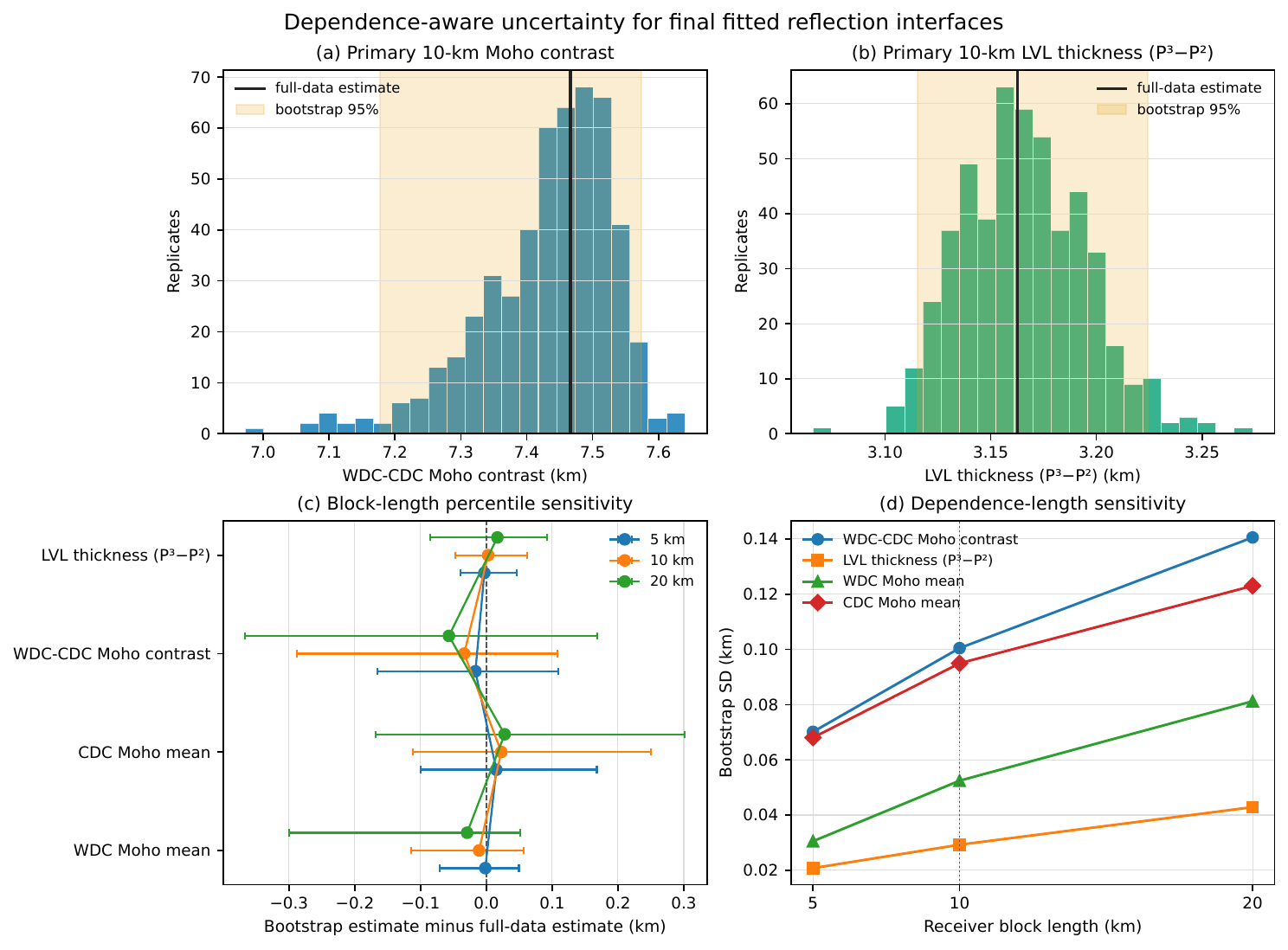}
\caption{Dependence-aware sampling uncertainty for the final fitted reflection
interfaces, conditional on the selected velocity, operator, regularisation and
seven shots. Receiver-coordinate blocks preserve every phase pick and its
multiplicity. The primary 10-km block lies nearest the 9.416-km residual
correlation scale and uses 500 stable replicates, of which 469 met a declared
solver convergence condition. (a,b) Bootstrap distributions and percentile
intervals for the Moho contrast and mean $P^{3}$--$P^{2}$ thickness.
(c,d) Sensitivity to 5-, 10- and 20-km blocks. At 10~km the raw contrast is
7.466~km, the bootstrap standard deviation 0.100~km and the percentile-95
interval 7.18--7.57~km. This scheme resamples \emph{within} shots; the
between-shot delete-one standard error is four times larger at 0.398~km.
Velocity and synthetic recovery bias are excluded.}
\label{fig:dependence}
\end{figure}

The empirical reflection-residual correlation length is 9.416~km, so the
primary non-circular moving-block bootstrap adopts 10-km receiver-coordinate
blocks inside each shot. All 500 primary replicates are formally stable and
preserve exact receiver and phase multiplicities, although 469 of the 500 met a
declared solver convergence condition and we record that figure rather than
describe all 500 as converged. The contrast has bootstrap standard deviation
0.100~km and percentile-95 limits of 7.18--7.57~km, with corresponding
full-profile LVL thickness limits of 3.115--3.224~km
(Fig.~\ref{fig:dependence}; Supplementary Fig.~S12). Contrast standard
deviations at block lengths of 5, 10 and 20~km are 0.070, 0.100 and 0.141~km.
We quote the percentile limits to two decimals because Monte Carlo error on a
95 per cent percentile from 500 replicates is itself of order 0.01--0.02~km.

The between-shot picture is markedly wider, and on the argument of this paper
it is the more relevant one. Deleting whole shots gives a standard error of
0.398~km, four times the within-shot bootstrap value, and omitting SP7 alone
moves the contrast to 7.008~km. Since the dependence we set out to respect
operates at gather level, a resampling scheme that reshuffles receivers inside
shots addresses a level below where the problem lives, and the between-shot
figure should be read as the honest sampling statement despite resting on only
seven clusters. Read as a jackknife, those same refits imply a bias of
$-0.480$~km, a bias-corrected contrast of 7.946~km and normal 95 per cent
limits of 6.686--8.247~km. With seven clusters this remains an influence
diagnostic rather than a primary interval, but it is notable that it agrees in
sign and approximate magnitude with the wholly independent synthetic estimate
below.

That targeted PmP experiment exposes a different effect. Across 15 transition
geometries and 50 noise realizations apiece, the recovered contrast carries a
mean design-class bias of $-0.361$~km, so adding 0.361~km to the raw estimate
gives 7.827~km. A pooled 95 per cent half-width of 0.381~km attains only 0.933
leave-one-scenario-out coverage, so we prefer the more conservative
scenario-envelope half-width of 0.454~km, whose minimum scenario coverage is
0.98. This calibration embraces acquisition geometry, regularisation,
parameterisation and independent pick noise within the tested transition class,
while excluding velocity and broader structural model classes. One limitation
deserves emphasis: the scenarios are built on plateau depths of 47.0 and
39.0~km, a contrast of 8.0~km, and therefore centre on the very model being
measured. The correction is consequently valid to the extent that the recovered
contrast is already approximately right, and it would need re-deriving were the
true contrast substantially different. The concordance with the independent
jackknife bias of $-0.480$~km is reassuring on that point but does not remove
the circularity.

\begin{figure}[H]
\centering
\includegraphics[width=0.98\textwidth,height=0.84\textheight,keepaspectratio]{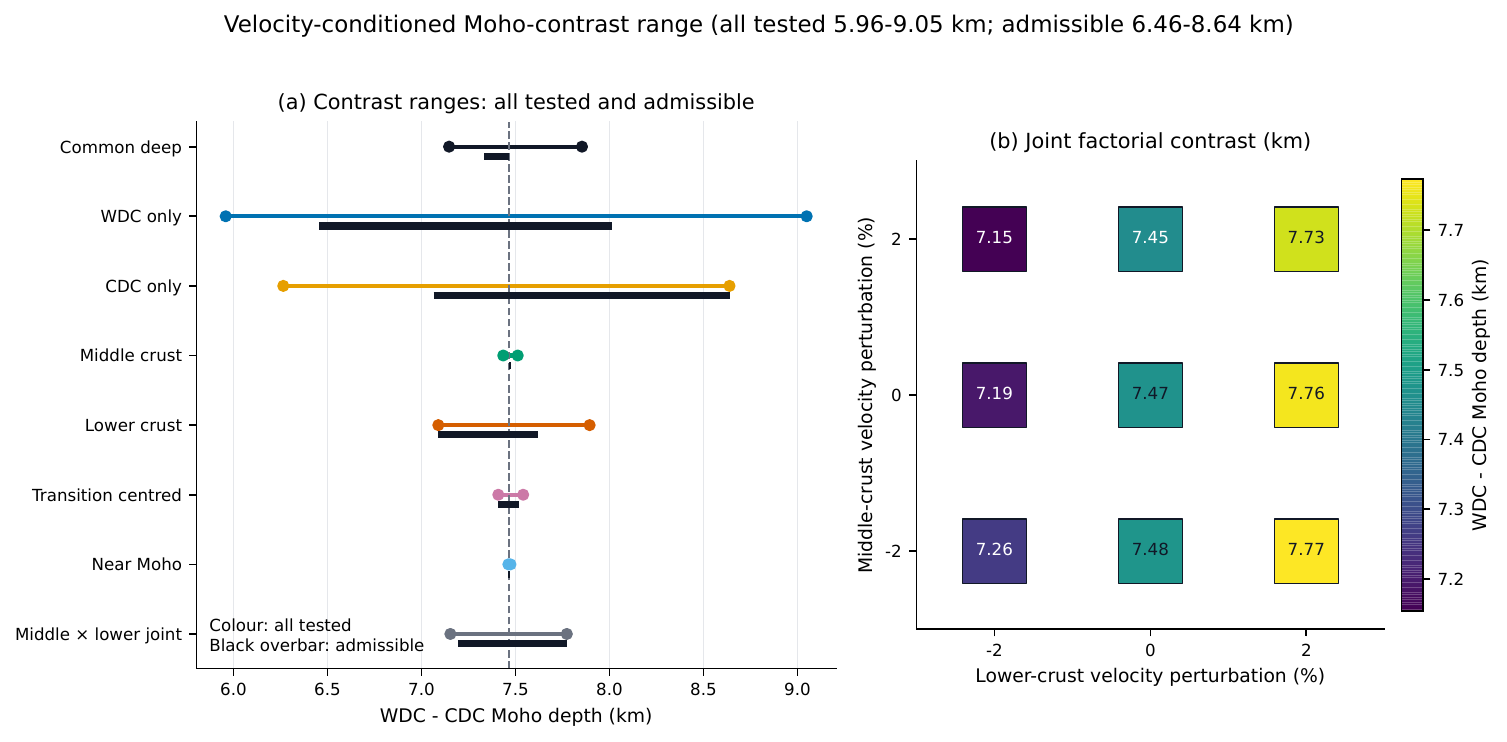}
\caption{Full nonlinear velocity--depth sensitivity of the exact-window
WDC--CDC Moho contrast, every unique perturbation rebuilding the fast-marching
table and refitting all six reflectors. (a) Coloured lines span all tested
amplitudes and black overbars retain the models passing the reflection,
interface-ordering, convergence and applicable refraction gates. (b) The true
$3\times3$ middle- and lower-crust factorial response. All tested raw contrasts
span 5.959--9.049~km and the admissible range is 6.457--8.639~km. Because the
$\pm3$ per cent amplitude is a declared design choice rather than a measured
bound, and because the gate compares an in-sample refit against a held-out
threshold, these are permissive design ranges and not probability intervals.}
\label{fig:veldepth}
\end{figure}

The nonlinear velocity ensemble is broader still (Fig.~\ref{fig:veldepth};
Supplementary Figs~S10 and S11). Of 58 ledger entries representing 47 unique
rebuilt fast-marching tables, 36 satisfy the predeclared reflection and
refraction gates. Changes common to the deep crust displace WDC and CDC largely
together, whereas province-localized changes do not cancel, and admissible raw
contrasts span 6.457--8.639~km. Every tested contrast stays positive, over
5.959--9.049~km, so a several-kilometre WDC--CDC difference is robust within
the tested ensemble even though its exact magnitude is not.

\subsection{Phase-specific evidence and trade-offs for the upper-crustal LVL}
\label{sec:lvlresults}

\begin{figure}[H]
\centering
\includegraphics[width=0.98\textwidth,height=0.84\textheight,keepaspectratio]{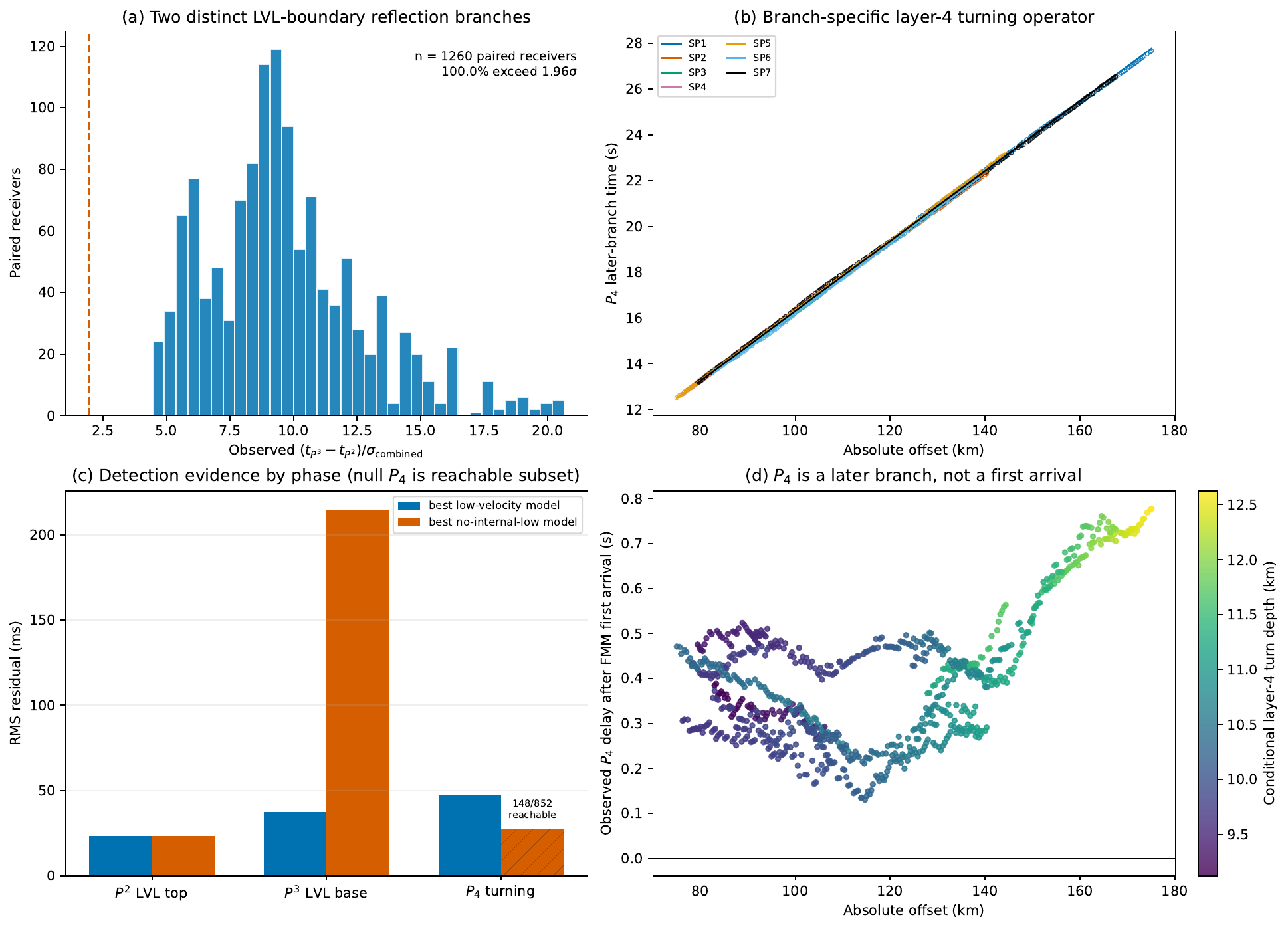}
\caption{Phase-specific evidence for the upper-crustal LVL. (a) Standardized
time separation between paired $P^{3}$ LVL-base and $P^{2}$ LVL-top reflections
at 1,260 common receivers, every separation being positive and exceeding 1.96
combined assigned standard deviations, with a minimum standardized value of
4.45. (b) Predictions of the separate layer-4 later-turning $P_4$ branch.
(c) Phase RMS for the selected low-velocity and continuous-null models, the
null giving a 214.8-ms $P^{3}$ RMS and reaching only 148 of 852 $P_4$
observations. (d) Every observed $P_4$ follows the fast-marching first arrival.
The $P_4$ result uses a corridor-averaged operator with one static per shot and
constitutes a bounded sensitivity test.}
\label{fig:lvldetect}
\end{figure}

The most direct observation is how far apart the two boundary reflections
arrive (Fig.~\ref{fig:lvldetect}). At 1,260 receiver coordinates both $P^{2}$
from the LVL top and $P^{3}$ from its base are recorded. In every single pair
$P^{3}$ is the later arrival; separations run from 0.249 to 1.154~s with a
median of 0.512~s, and each exceeds 1.96 times its combined assigned
uncertainty, the smallest standardized separation being 4.45. This establishes
two distinct reflected branches without extrapolating any turning phase, but it
does not by itself identify a unique velocity function between them, and it
should not be read as establishing a low-velocity interval on its own.

\begin{figure}[H]
\centering
\includegraphics[width=0.98\textwidth,height=0.84\textheight,keepaspectratio]{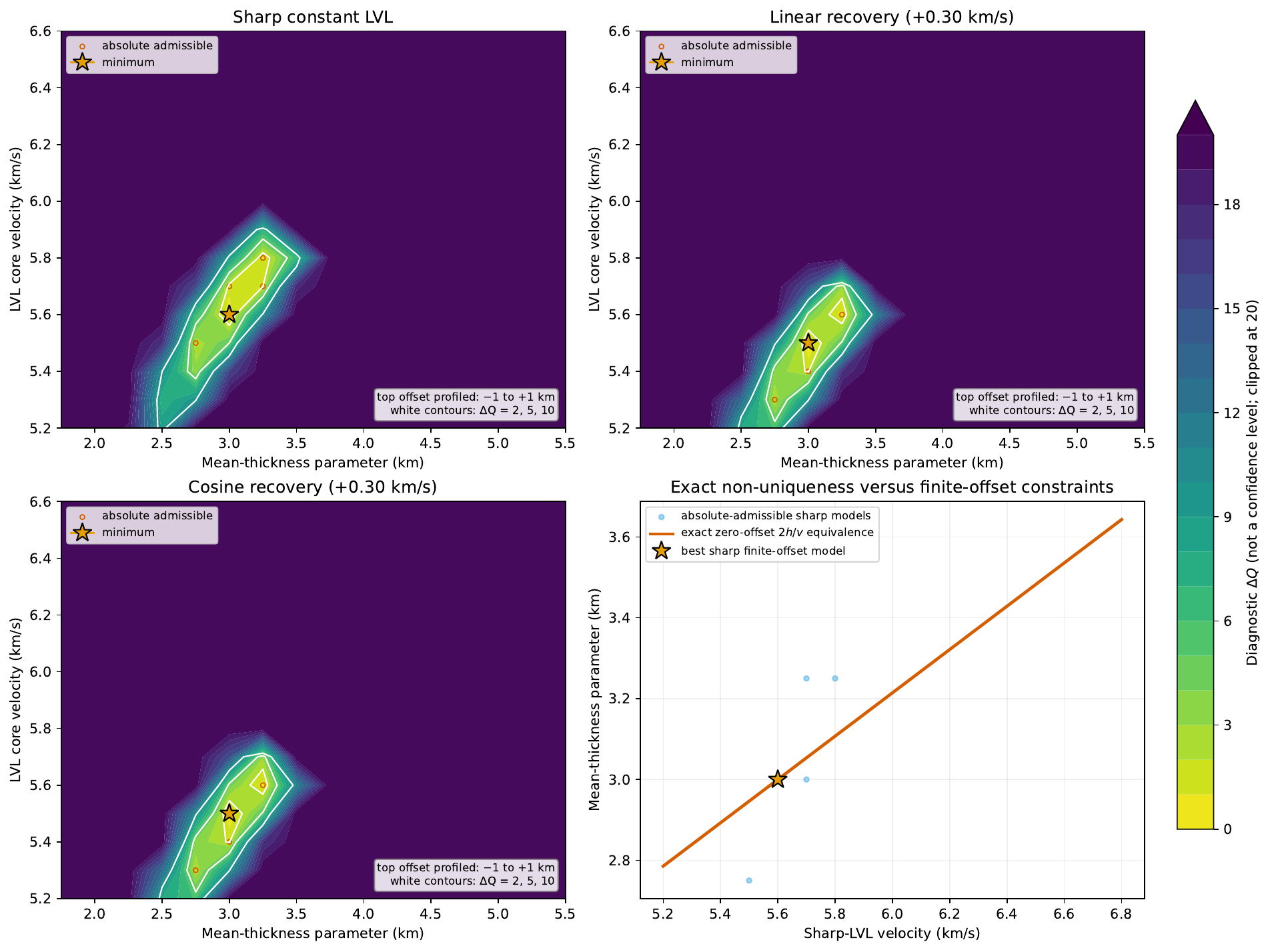}
\caption{Conditional LVL thickness--velocity--shape trade-off across 6,624
predeclared candidates. Panels show sharp, linear-recovery and cosine-smoothed
profiles after profiling over top offset, open symbols marking the 13
candidates that pass the absolute $P^{2}$, $P^{3}$ and conditional $P_4$ gates.
The lower-right panel contrasts exact zero-offset $2h/v$ equivalence with
finite-offset constraints. White $\Delta Q$ contours are diagnostic score
differences and not confidence regions. The preferred local model carries a
3.0-km mean-thickness scan parameter, referenced to the 3.224-km nodal mean,
and a 5.5-km\,s$^{-1}$ core velocity, but internal shape remains unresolved.}
\label{fig:lvltrade}
\end{figure}

Taken over the whole profile, the independently fitted six-interface solution places the exact mean $P^{3}$--$P^{2}$ separation at 3.163~km, the conditional 10-km block bootstrap bracketing it between 3.115 and 3.224~km at 95 per cent. The subsequent
phase-specific physical scan contains 13 rule-based admissible low-velocity
candidates (Fig.~\ref{fig:lvltrade}). All retain the final LVL-top position on
the tested 0.25-km top-offset grid, their mean-thickness scan parameters
running 2.75--3.25~km and their core velocities 5.3--5.8~km\,s$^{-1}$; recall
that these scan parameters are referenced to the 3.224-km nodal mean and are
not the same quantity as the 3.163-km window integral. The minimum-score model
combines a linear 0.30-km\,s$^{-1}$ internal recovery, zero top offset, a
3.00-km mean-thickness parameter and a 5.50-km\,s$^{-1}$ core velocity, giving
mean top and base depths of 6.684 and 9.684~km and a node-wise thickness of
1.955--4.509~km. Because sharp, linear-recovery and cosine-smoothed candidates
all satisfy the admissibility rules, the internal shape remains unresolved.

\begin{table}[H]
\centering
\caption{Phase-specific LVL metrics. The $P_4$ values use a
shot-static-corrected corridor-averaged turning operator. The continuous-null
$P_4$ RMS covers only its 148 reachable picks and is therefore not comparable
with the 852-pick value for the selected model. Deeper-phase RMS values are
listed to show that the LVL perturbation, which the admissibility rules do not
gate on these phases, nevertheless alters their predictions; the corresponding
freely fitted geometry values are 60.113, 63.957, 43.464 and 57.678~ms.}
\label{tab:lvl}
\small
\setlength{\tabcolsep}{4pt}
\begin{tabular}{@{}lrrr@{}}
\toprule
Model & $P^{2}$ RMS / $n$ & $P^{3}$ RMS / $n$
& $P_4$ conditional RMS / reachable $n$ \\
\midrule
Selected low velocity & 23.448 / 1486 & 37.237 / 1528
& 47.608 / 852 of 852 \\
Best continuous null & 23.448 / 1486 & 214.812 / 1528
& 27.617 / 148 of 852 \\
\midrule
\multicolumn{4}{@{}l}{\emph{Selected model, deeper phases (ungated):}
$P^{4}$ 60.458, $P^{5}$ 63.481, $P^{6}$ 48.768, PmP 58.295~ms} \\
\bottomrule
\end{tabular}
\end{table}

For the selected model all 852 $P_4$ offsets are reachable, every conditional
turning point falls inside layer 4, and the standardized MSE is 0.907. Every
observed $P_4$ arrives 0.130--0.778~s after the corresponding fast-marching
first arrival, which confirms it to be a later branch. The continuous null,
whose own best configuration sits at a 4.5-km mean thickness rather than being
frozen at the LVL geometry, permits only 17.4 per cent of the observed $P_4$
offsets and returns a $P^{3}$ RMS of 214.812~ms with standardized MSE 18.458.
Introducing the 112 ambiguous $P^{2}/P_2$ records by way of a separate sensitivity calculation alters none of the selected top offset, thickness, core velocity or shape. Because one nuisance static is fitted per shot and the $P_4$
operator is one-dimensional along each corridor, its 47.608-ms RMS constitutes
conditional branch evidence rather than an independently calibrated
absolute-time prediction. The adopted $P_1$--$P_2$ subset stays insensitive to
the LVL, and that statement must not be carried over to $P_4$.

Two qualifications on the scan follow from Table~\ref{tab:lvl}. First, the
selected LVL improves $P^{3}$ from the 43.140~ms of the freely fitted geometry
to 37.237~ms, which happens because the scan adjusts the mean thickness
governing the $P^{3}$ interface and so is expected rather than anomalous.
Second, since the LVL overlies every deeper reflector, altering its velocity
shifts all deeper reflection times: relative to the geometry fit, $P^{6}$
degrades from 43.464 to 48.768~ms with a bias of $-21.27$~ms and PmP moves from
57.678 to 58.295~ms with a bias of $-11.48$~ms. The latter corresponds to
roughly 35~m of Moho depth. The effect is small but it is not zero, so the LVL
parameters and the Moho estimate are coupled and our admissibility rules,
gating only $P^{2}$, $P^{3}$ and $P_4$, do not police it. Finally, although the
null is profiled over the scan grid rather than pinned to the LVL solution,
both it and the low-velocity candidates inherit a frozen six-interface lateral
geometry that was itself fitted with a low-velocity interval present; a fully
independent null would re-invert that geometry from the start.

\subsection{Upper-crustal velocity}

\begin{figure}[H]
\centering
\includegraphics[width=0.98\textwidth,height=0.84\textheight,keepaspectratio]{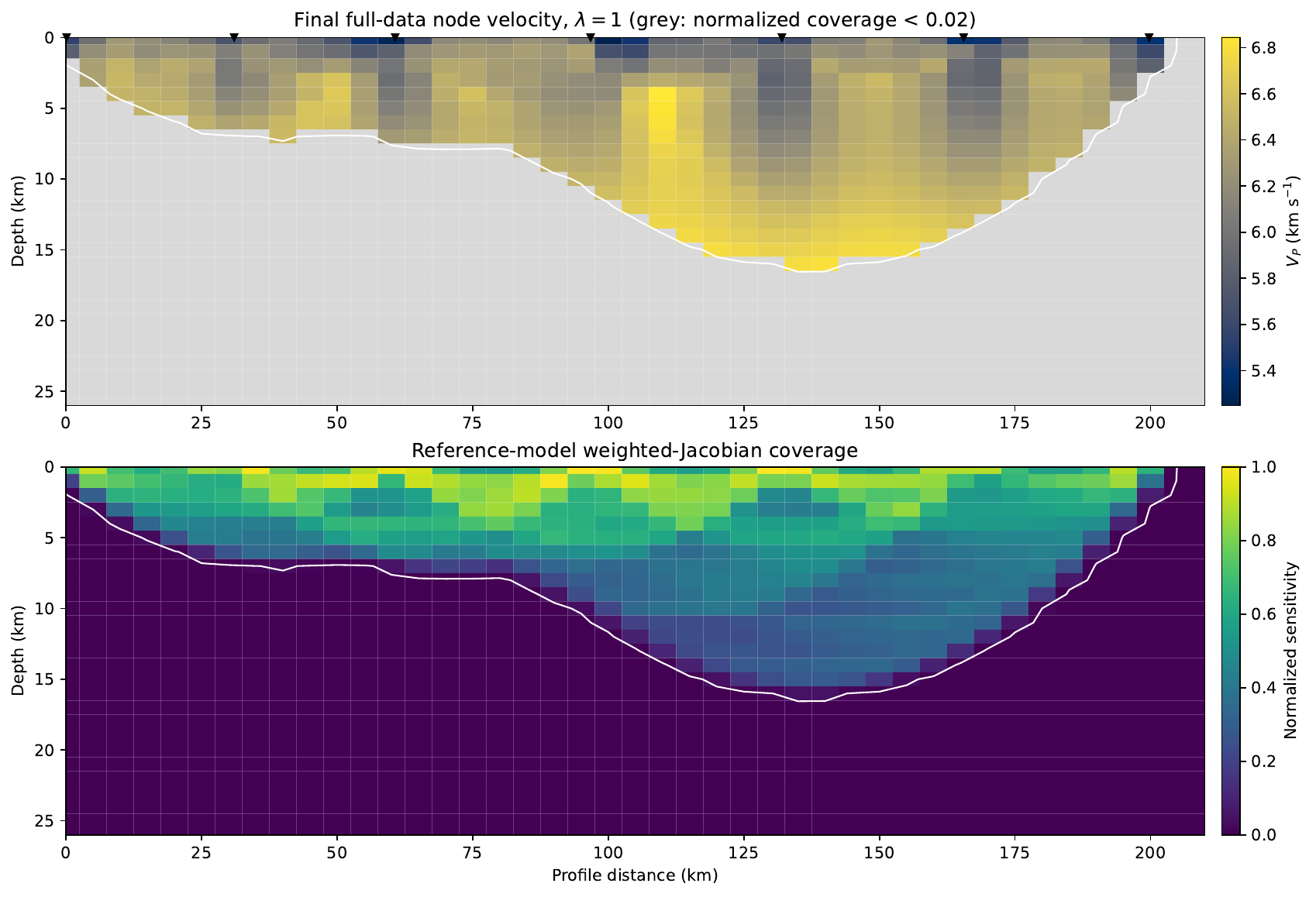}
\caption{Final full-data $P_1$--$P_2$ velocity model selected at $\lambda=1$
(top) with normalized reference-model weighted-Jacobian coverage (bottom). Grey
cells fall below 0.02 coverage, the white line marks that support boundary and
triangles mark shots. The 2,000 primary first arrivals constrain the upper crust
only. Although the deepest traced ray reaches 20.405~km, structure outside the
displayed support is not interpreted as a refraction result.}
\label{fig:refr}
\end{figure}

Tuned separately to $\lambda=1$, the all-data inversion reaches 16.57~ms RMS over the 2,000 $P_1$--$P_2$ picks, equivalent to $\chi^2/N=0.439$
(Fig.~\ref{fig:refr}). A reduced chi-squared well below unity, obtained with
1,161 unknowns against 2,000 observations, indicates that the model is fitting
more closely than the noise level warrants. Set beside the 102.60-ms nested
outer residual, the small in-sample figure is evidence of interpolation
capacity rather than of complete-shot predictive calibration.

Only 415 of the 1,161 velocity nodes attain normalized reference-model
weighted-Jacobian coverage of at least 0.02. Inside that gate $V_P$ spans
5.25--6.84~km\,s$^{-1}$, and the gate reaches 16~km only between 135 and
140~km profile distance. Figure~\ref{fig:refr} greys every node beneath the
threshold and no velocity below the white boundary is interpreted. That some
nodes outside the support mask run up against the imposed 7.2-km\,s$^{-1}$
ceiling reinforces the exclusion. The recovered field is smooth and free of
imposed layers, but because the adopted first arrivals do not sample the LVL
that smoothness must not be read as evidence against the layer constrained by
the reflected and later-branch phases.

\subsection{Targeted Moho recovery}

Our final synthetic experiment uses the actual 1,494-pick PmP geometry and
tests 15 noise-free transitions, pairing centres at 100, 110 and 120~km with
sharp, 20, 40, 60 and 100-km-wide forms. Fifty independent pick-noise
realizations follow each, giving 750 noisy inversions. All 750 solutions are
formally stable, 745 satisfy the primary convergence declaration, and the five
continuations alter neither a node nor a province metric at the reported
precision.

\begin{figure}[H]
\centering
\includegraphics[width=0.98\textwidth,height=0.84\textheight,keepaspectratio]{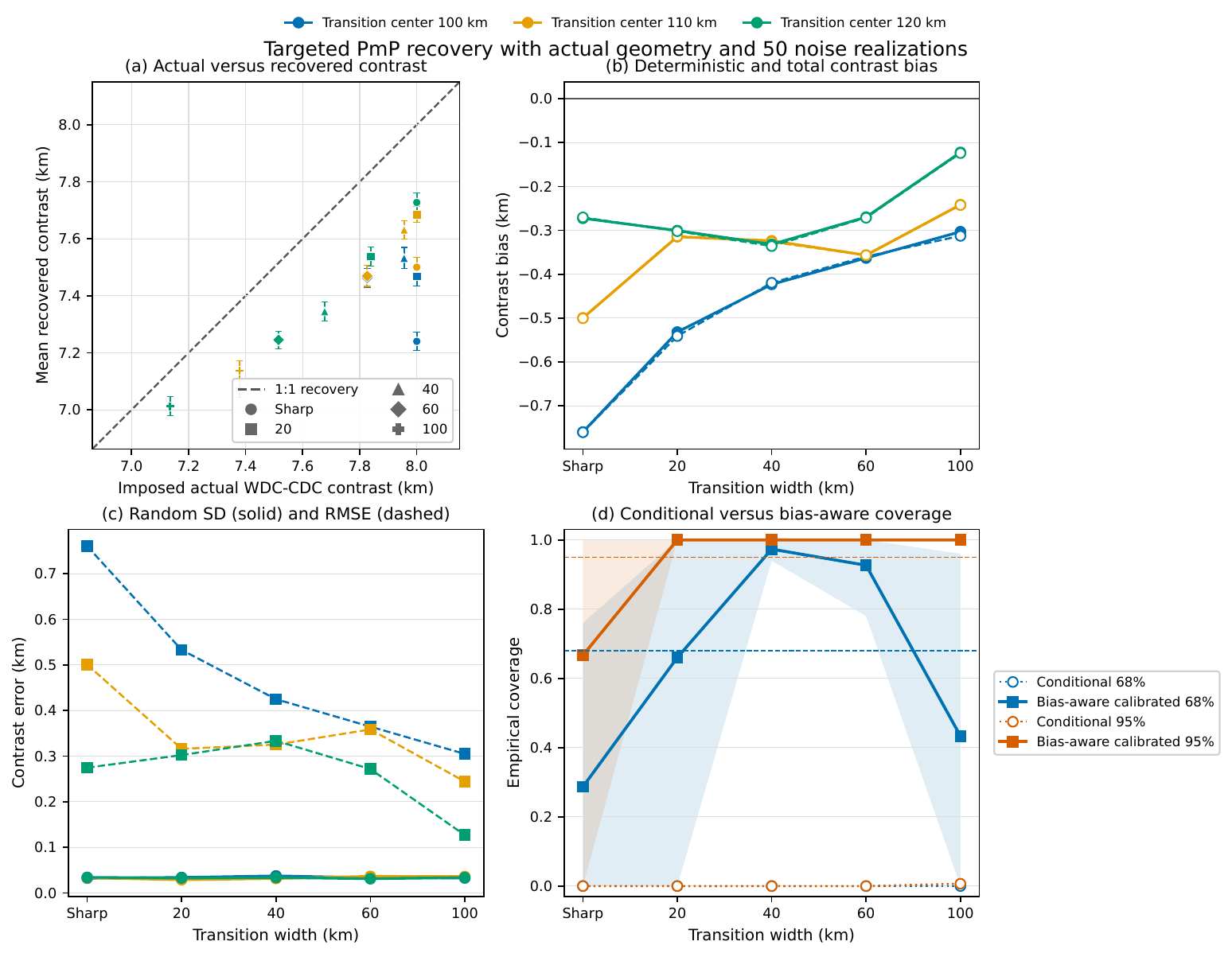}
\caption{Targeted PmP recovery using the actual 1,494-pick geometry with 50
independent noise realizations for each of 15 transition scenarios.
(a) Imposed against ensemble-mean recovered WDC--CDC contrast.
(b) Deterministic and total recovery bias. (c) Random standard deviation and
RMSE. (d) Conditional and leave-one-transition-scenario-out bias-aware
coverage. Mean design-class bias is $-0.361$~km, so the prespecified correction
carries the raw 7.466-km contrast to 7.827~km, and the conservative conditional
scenario-envelope 95 per cent half-width is 0.454~km. Because the scenarios are
built on plateau depths of 47.0 and 39.0~km they centre on the measured model,
so the correction presupposes an approximately correct contrast. Velocity, shot
correlation and broader model-class uncertainty are excluded and this is not a
total interval.}
\label{fig:resmoho}
\end{figure}

Imposed exact-window contrasts span 7.135--8.000~km. Across scenarios the
deterministic recovery bias runs from $-0.760$ to $-0.124$~km while the
within-scenario random standard deviation is merely 0.0289--0.0374~km
(Fig.~\ref{fig:resmoho}; Supplementary Fig.~S13). Acquisition geometry,
regularisation and parameterisation bias therefore dominate independent pick
noise for this estimand, and conditional local intervals accordingly show poor
uncorrected coverage. Leave-one-transition-scenario-out calibration is what
motivates the bias correction and conservative design envelope of
Table~\ref{tab:budget}. Shifting the inner province-window limits 10~km outward leaves a conservative synthetic-only 95 per cent half-width of 0.264~km, whereas shifting them 10~km inward inflates it to 0.864~km (Supplementary Fig.~S14), so
province definitions form part of the estimand rather than being an innocuous
plotting choice.

\section{Discussion}

\subsection{What may and may not be claimed}
\label{sec:claims}

A number of statements survive within their declared scope. The permanent phase
ledger separates all six reflected branches, ray 2.2 ($P^{2}$, LVL top) and ray
3.2 ($P^{3}$, LVL base) included, from the later-turning ray 4.1 ($P_4$). Each
reflected phase prefers its own boundary by a margin far exceeding pick
uncertainty, the fixed-velocity geometry is insensitive to both tested starting
models, and the positive WDC--CDC Moho contrast persists across every velocity
field we tried. Together these support a thicker crust beneath WDC than beneath
CDC along this profile.

The magnitude, though, does not warrant a single small error bar. The
fixed-velocity raw contrast, the dependence-aware interval, the synthetic
bias-adjusted contrast and the admissible velocity range answer different
questions. Perturbations common to the deep crust displace both provinces
together whereas province-specific ones change their difference substantially,
so the contrast benefits from common-mode cancellation without being immune to
velocity--depth ambiguity. Nor does the targeted synthetic design estimate a
transition-centre parameter; it calibrates province averages over declared
windows. Any spatial coincidence between the fitted Moho change and the surface
trace of the Chitradurga Shear Zone is therefore a hypothesis-level
consistency, not a calibrated localization and not evidence of a sharp Moho
step.

Three limitations bear more weight than the rest and we would rather state them
than let them be inferred. The first concerns velocity. Nothing in this dataset
constrains crustal velocity below roughly 16~km, yet the Moho lies between 39
and 47~km, so the depth conversion at Moho level rests entirely on the field
inherited from \citet{beherakumar2022}. Our $\pm3$ per cent perturbation
envelope is a declared design choice and not a measured bound; since depth
scales roughly with velocity, it translates to about $\pm1.4$~km on a 46.9-km
Moho, which is close to what the admissible range shows. Were the defensible
spread for Archean lower-crustal velocity appreciably wider, this axis would
grow proportionately and would dominate every other entry in
Table~\ref{tab:budget}. Establishing that spread independently, whether from
petrophysical constraints or from joint inversion with gravity in the manner of
\citet{nielsen2000}, is the single most valuable extension of this work. The
second concerns the refraction result, which as argued in
Section~\ref{sec:calib} reflects aperture extrapolation and an unrepresentable
near-surface at least as much as pick-error miscalibration; a parameterisation
admitting low near-surface velocities, or explicit shot statics, would be
needed to separate them. The third is that the synthetic design class is
centred on the measured model, so its bias correction presupposes an
approximately correct answer.

The LVL result likewise separates detection from parameter resolution. Matched
$P^{2}/P^{3}$ observations establish two distinct reflected branches, and the
poor $P^{3}$ fit together with the severe $P_4$ reachability failure of the
continuous null support an internal low-velocity interval within the tested
model class. They do not fix its boundary sharpness or its velocity profile.
This reading is consistent with, while deliberately weaker than, the
geological interpretation of \citet{kumarbehera2023}, who read the same
interval as an eastward-dipping detachment separating the Dharwar schist
belts from the underlying gneisses: our analysis supports the existence and
mean geometry of the interval but shows that its internal velocity structure
is not resolved by these phases.
The exact sharp-layer relation $\tau_{\rm LVL}=2h/v$ leaves thickness and
velocity non-unique, so the admissible grid ranges are sensitivity bounds and
not confidence intervals. Refraction velocities, finally, are interpreted only
inside the spatially varying 0.02 coverage mask; one deep ray reaching 20~km is
no licence to interpret the whole model at that depth.

\subsection{Reconciling published crustal thicknesses}

Published estimates do not form a single like-for-like depth range
(Supplementary Table~S1). \citet{rao2015b} obtain roughly 42 and 38~km from a
wide-angle model across the WDC--EDC convergence zone; \citet{mall2012} report
45 and 39~km from a six-layer seismic model; \citet{julia2009} derive 45--50~km
beneath WDC and 32--35~km beneath EDC from joint receiver-function and
surface-wave inversion; and \citet{gupta2005} reach comparable conclusions from
teleseismic waveform modelling. The earlier ray-trace interpretation of this
same profile by \citet{beherakumar2022} and \citet{kumar2022thesis} gives about
49 and 38~km. Reported uncertainties, spatial footprints and operational
definitions of the Moho are not uniform across these studies, and a
terminological caution applies to the comparison itself: we label the
north-eastern block CDC, whereas most of this literature divides the craton
into western and eastern domains, so our CDC window and a published EDC value
need not describe the same crust.

Our own numbers are 46.915~km for the fixed-velocity WDC mean, moving to
47.067~km under the synthetic design correction with a conditional
design-envelope half-width of 0.446~km, while admissible nonlinear velocity
models span 45.763--47.500~km. These ranges do not encompass all published
values, but as they are not a complete total uncertainty they cannot assign the
residual literature spread to any single cause. Differing sampling locations,
phase selections, velocity parameterisations, three-dimensional structure and
method-specific Moho definitions all remain plausible contributors, much as
\citet{zelt1999} anticipated.

One comparison deserves separate treatment because it is internal to our own
work. \citet{beherakumar2022} reported approximately 49 and 38~km from these
very data, a contrast near 11~km, against the 7.466~km we obtain here. Nothing
in the observations changed; the revision follows from how the reported
quantity is defined and fitted. Where the earlier study quoted depths at
interpreted positions in a ray-trace model, we integrate a linearly
interpolated reflector exactly across declared province windows, and the
regularised simultaneous six-interface fit smooths the transition that a
layer-by-layer interpretation renders more abruptly. The two figures are
therefore answers to different questions rather than a disagreement about the
crust, which is precisely the argument of this paper applied to itself: a
several-kilometre change in a headline contrast can follow from estimand
definition alone, with no new data and no error in either analysis.

The same caution attends any preference for contrasts over absolute depths. Our
raw contrast is 7.466~km and the synthetic design-class correction gives
7.827~km, yet the admissible velocity range is 6.457--8.639~km. A differential
quantity is most valuable when the dominant perturbation is common to both
windows and is considerably less protected against province-specific model
error. Cross-study comparisons should accordingly state averaging windows,
phase set and velocity family alongside both absolute depths and their
contrast, a discipline equally relevant wherever independent methods disagree
by kilometres in other cratons, as \citet{jull2001}, \citet{reading2003} and
\citet{behera2006} have each documented.

\subsection{Implications for reporting practice}

Four practices follow from this exercise. First, observations should carry
permanent identifiers through phase reconciliation, so that restored,
ambiguous, conditional and excluded records can never be pooled silently.
Second, assigned pick errors and regularisation ought to be tested by grouped
prediction with every selection step confined to training shots, and the
grouping should be checked for exchangeability before pooled predictive error
is interpreted as a calibration scale. Third, the reported quantity must be
defined mathematically, because exact profile-window integrals and simple node
means are not interchangeable, as the difference between our 7.466~km and
7.457~km illustrates at small scale and the comparison with our own earlier
work illustrates at large. Fourth, conditional resampling, shot leverage,
synthetic recovery bias, velocity alternatives, boundary choices and numerical
error each belong in separate rows with separate scopes. Here 36 of 58
predeclared velocity-ledger entries pass the final gates, and averaging
accepted against rejected models would obscure uncertainty rather than
quantify it. We would add a fifth: where resampling can act at more than one
level, the level matching the dependence being claimed should be the one
reported foremost, which is why we place the between-shot spread of 0.398~km
ahead of the within-shot 0.100~km.

\section{Conclusions}

Auditing the Perur--Chikmagalur dataset yields 2,000 primary $P_1$--$P_2$ first arrivals alongside 9,919 primary reflections. We identify rays 2.2 and 3.2 as two separate shallow reflections, $P^{2}$ returning from the LVL top and $P^{3}$ from its base, whereas the 852 records carried under code 3 constitute the later-turning $P_4$ branch and feed a labelled conditional analysis only.

Complete-shot prediction rather than all-data residual sets the calibration
scale, and it must be read phase by phase. Reflections pool to 68-ms held-out
RMS with standardized MSE 2.07 and 0.85 nominal-95 coverage, but the assigned
uncertainties are close to adequate for $P^{3}$ and $P^{6}$ and understated
threefold for $P^{4}$ and $P^{5}$; for PmP, which carries every Moho claim,
the held-out figures are 73~ms and 2.12. The refraction inversion fits all data
at 17~ms RMS yet predicts untouched shots at 103~ms with 0.53 nominal-95
coverage, a failure driven by extrapolation beyond the sampled aperture and by
a velocity parameterisation unable to represent near-surface weathered
material, and only partly by dependence among picks.

At fixed velocity the exact WDC and CDC Moho means are 46.9 and 39.4~km, with a
raw contrast of 7.47~km. Resampling receivers within shots gives a conditional
95 per cent contrast interval of 7.18--7.57~km, but deleting complete shots
gives a standard error four times larger, and we regard that between-shot
figure as the honest sampling statement. Two independent experiments concur
that the raw contrast is biased low, by 0.36~km from targeted synthetic
recovery within the tested transition class and by 0.48~km from the delete-one
jackknife, giving bias-adjusted contrasts of 7.83 and 7.95~km. Admissible
nonlinear velocity models span 6.46--8.64~km, and because this ensemble rests
on a declared $\pm3$ per cent design choice rather than a measured bound it is
the axis most in need of independent constraint. No single total interval is
claimed.

Across 1,260 matched receivers $P^{3}$ trails $P^{2}$ without exception, and not one separation falls within 1.96 combined assigned standard deviations of zero. Mean-thickness parameters surviving the phase-specific scan run from 2.75 to 3.25~km, with core velocities between 5.3 and 5.8~km\,s$^{-1}$; the selected model reaches all 852 $P_4$
offsets whereas the continuous null reaches only 148 and yields a 214.8-ms
$P^{3}$ RMS. The LVL is supported within this model class, though its internal
shape is unresolved and its parameters are mildly coupled to the deeper
interfaces. More generally, a small conditional sampling spread on a fitted
interface must never be mistaken for uncertainty in velocity, geometry,
estimand definition or model class.

\section*{Acknowledgements}

We thank CSIR-NGRI for the 3-C wide-angle seismic data. Authors take full responsibility for the published content.
\section*{Author contributions}

\textbf{Deepak Kumar}: conceptualisation, methodology, software, formal
analysis, investigation, visualisation, writing of the original draft.
\textbf{Laxmidhar Behera}: resources, data curation, validation, supervision,
review and editing.
\textbf{Wojciech Czuba}: supervision, review and editing.

\section*{Declarations}

\paragraph{Funding.} The authors received no funding for this work.

\paragraph{Competing interests.} The authors have no competing interests to
declare.

\paragraph{Data availability.} CSIR-NGRI acquired  3-C wide angle seismic data  which is used in this study; access to them may be requested from CSIR-NGRI. Upon
acceptance we shall lodge the analysis code in a public archive under a DOI,
covering the fast-marching forward solver, the reciprocity-based reflection
table, the calibration protocol and the figure scripts, along with the
machine-readable result ledgers from which every number quoted above is
drawn.


\clearpage
\section*{Figures}

\clearpage

\clearpage

\clearpage

\clearpage

\clearpage

\clearpage

\clearpage

\clearpage

\clearpage

\clearpage

\clearpage

\clearpage

\clearpage
\graphicspath{{supplement/figures/}{}}
\setcounter{figure}{0}
\renewcommand{\thefigure}{S\arabic{figure}}
\setcounter{table}{0}
\renewcommand{\thetable}{S\arabic{table}}

\section*{Supplementary images}

\begin{figure}[p]
\centering
\includegraphics[width=\textwidth,height=0.72\textheight,keepaspectratio]
  {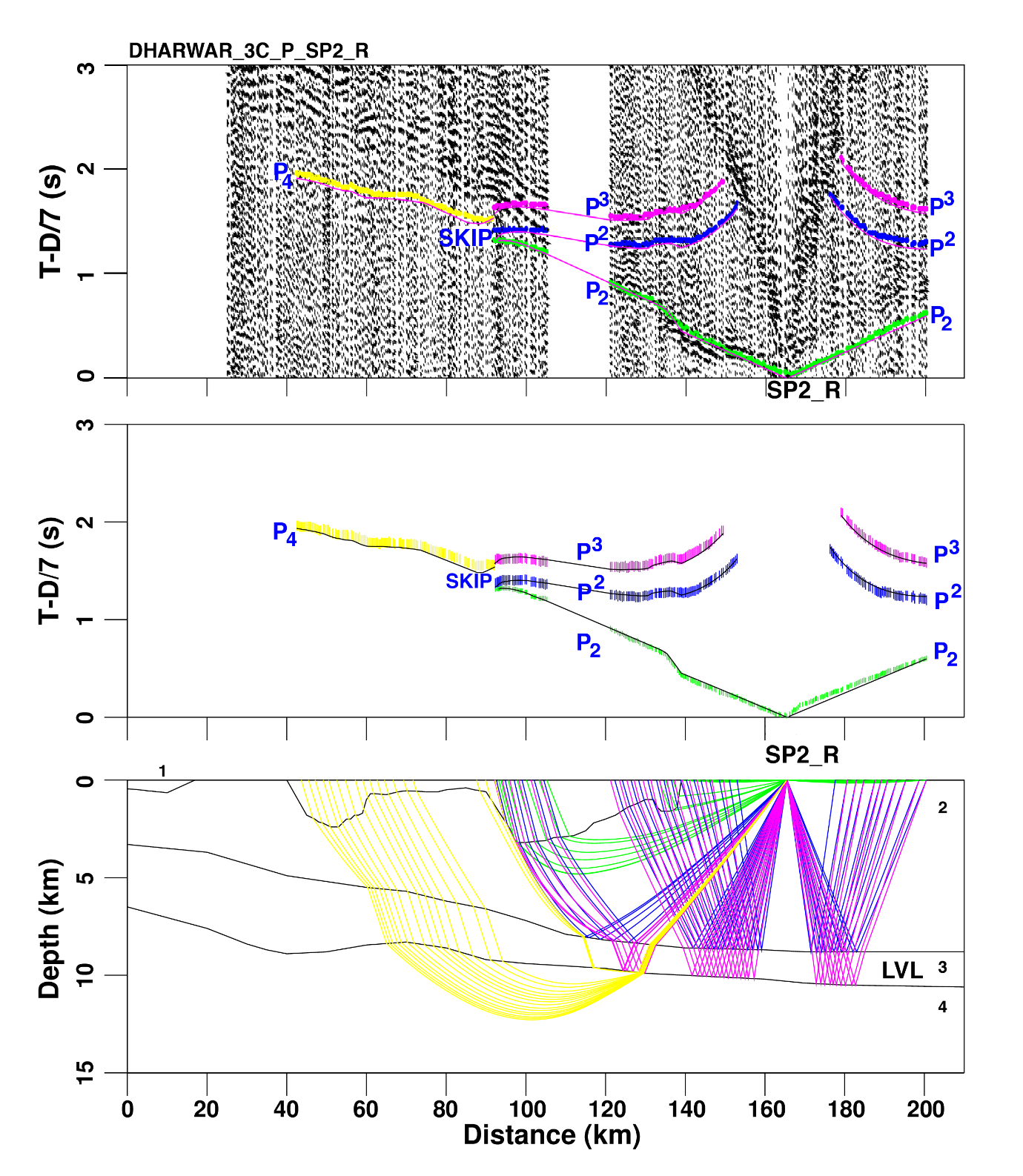}
\caption{Inherited SP2 upper-crustal radial-component record section (top),
legacy calculated-response/pick comparison (middle), and legacy ray diagram
(bottom), after Kumar (2022) and Behera and Kumar (2022). The corrected phase
interpretation is explicit: blue \(P^{2}\) is \textsc{rayinvr} ray 2.2,
reflected from the bottom of layer 2 and hence the \textbf{top of the LVL};
magenta \(P^{3}\) is ray 3.2, reflected from the bottom of layer 3 and hence
the \textbf{base/bottom of the LVL}. Green \(P_2\) and yellow \(P_4\) are
turning phases. The labelled SKIP interval denotes the missing \(P_3\) turning
branch through the LVL, whereas \(P_4\) turns below it. This panel is inherited
observational and historical ray-tracing context, not an output of the present
inversion, validation, or uncertainty analyses.}
\label{fig:supp-sp2}
\end{figure}
\clearpage
\begin{figure}[p]
\centering
\includegraphics[width=\textwidth,height=0.77\textheight,keepaspectratio]
  {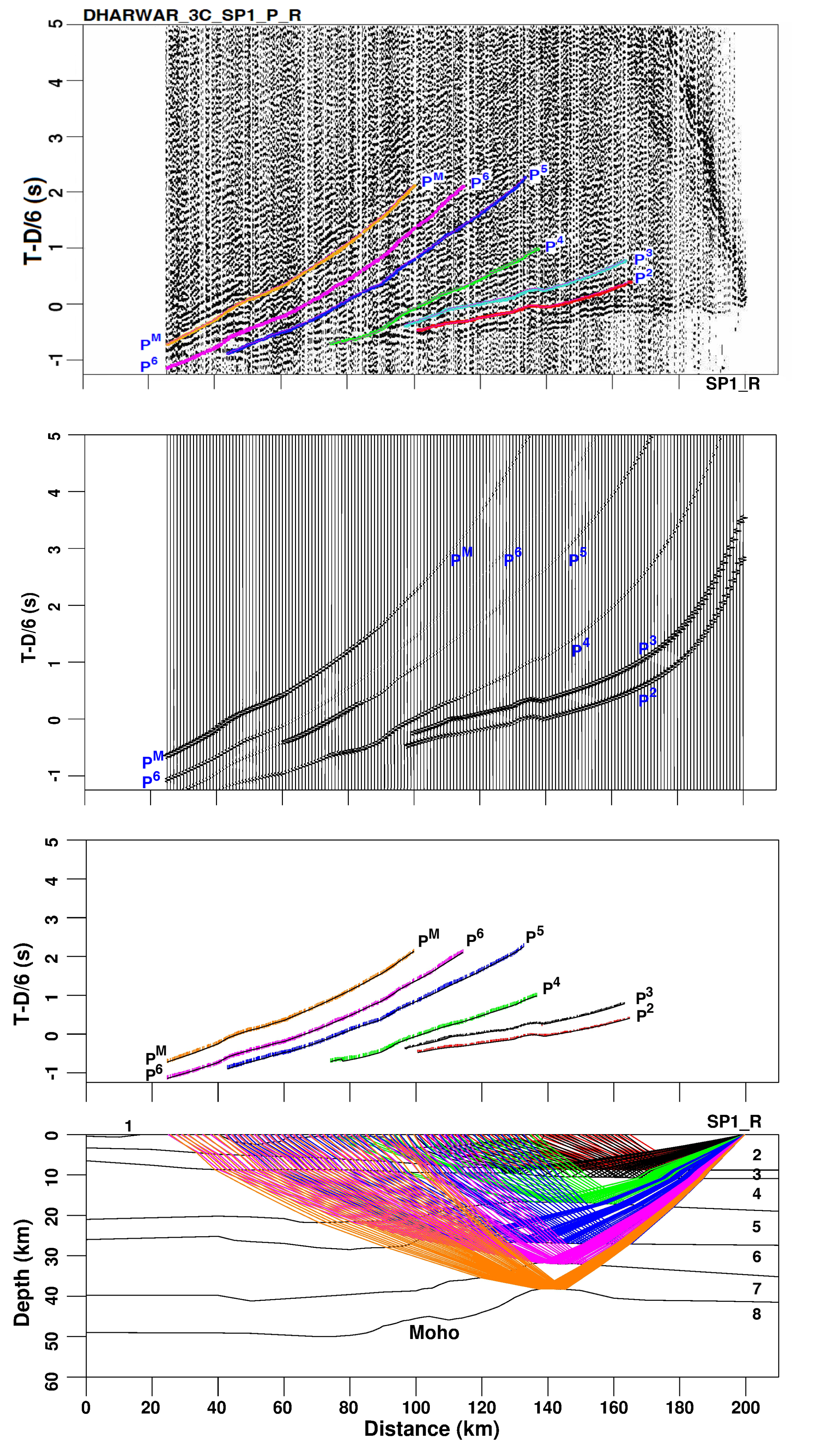}
\caption{Inherited full-crustal SP1 radial-component observed gather with
phase picks (top), legacy calculated response (second panel), traveltime
overlay (third panel), and traced rays and interfaces (bottom), after Kumar
(2022) and Behera and Kumar (2022). Labels are read under the permanent audit:
\(P^{2}\) is ray 2.2 reflected from the LVL top, \(P^{3}\) is ray 3.2
reflected from the LVL base, and the deeper \(P^{4}\), \(P^{5}\), \(P^{6}\),
and legacy \(P^{M}\) branches are reflections from successively deeper
boundaries (legacy \(P^{M}\) is PmP in the revised manuscript). Superscripts
must not be confused with subscript turning phases. This inherited panel
documents observational and historical ray-tracing context only; it is not
the new inversion, validation, or uncertainty result.}
\label{fig:supp-sp1}
\end{figure}
\clearpage
\begin{figure}[p]
\centering
\includegraphics[width=0.99\textwidth,height=0.70\textheight,keepaspectratio]
  {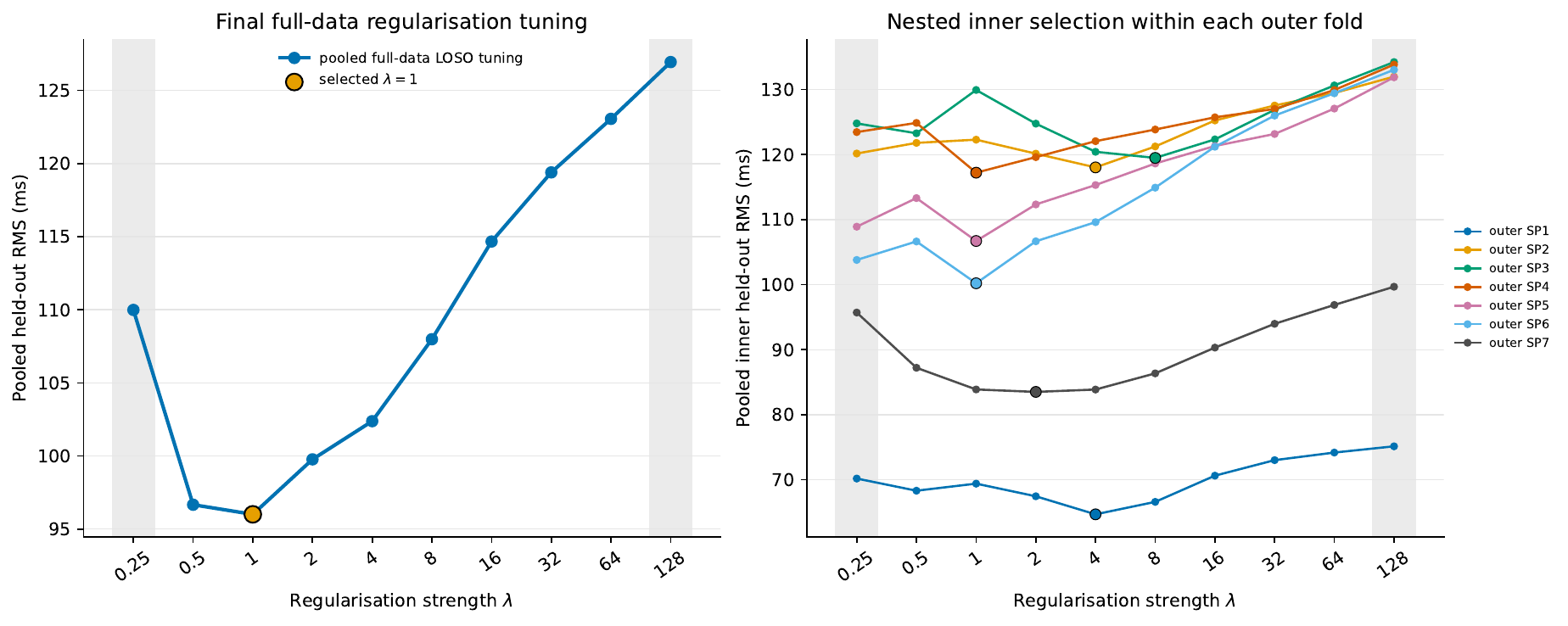}
\caption{First-arrival regularisation selection. The full-data
leave-one-complete-shot-out tuning pass selects \(\lambda=1\); nested
training-only tuning within outer SP1--SP7 selects
\(\lambda=4,4,8,1,1,1,\) and \(2\), respectively. Every selected value is
interior to the searched \(0.25\)--\(128\) grid. The full-data tuning
predictions select the final fit and are not substituted for the untouched
nested outer predictions used to report performance.}
\label{fig:supp-refraction-lambda}
\end{figure}
\clearpage
\begin{figure}[p]
\centering
\includegraphics[width=0.99\textwidth,height=0.70\textheight,keepaspectratio]
  {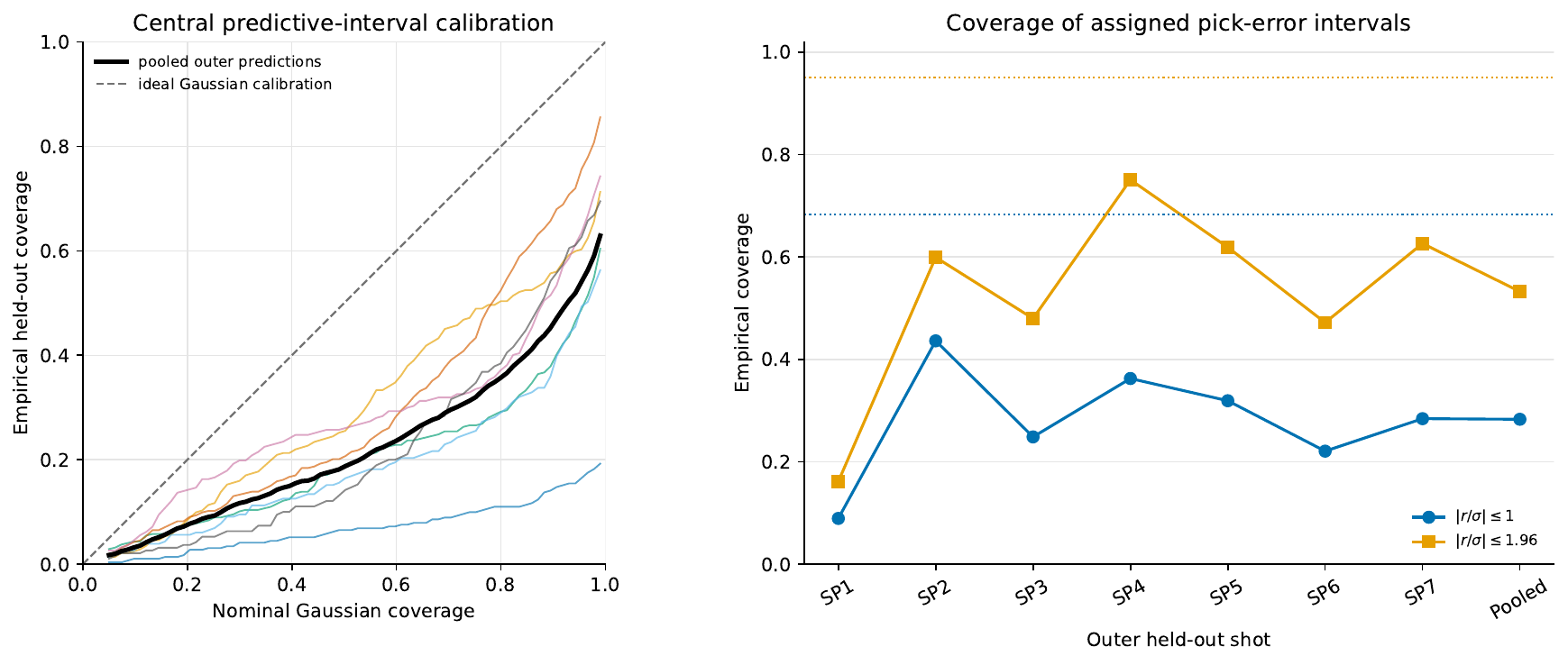}
\caption{Predictive calibration of the untouched outer \(P_1\)--\(P_2\)
predictions. The central calibration curve and complete-shot coverages are
evaluated against the original assigned 25-ms pick uncertainty. Pooled
empirical coverage is 0.283 for \(\lvert r/\sigma\rvert\leq1\) and 0.5325 for
\(\lvert r/\sigma\rvert\leq1.96\); the assigned pick-error intervals therefore
substantially under-cover complete-shot prediction error.}
\label{fig:supp-refraction-calibration}
\end{figure}
\clearpage
\begin{figure}[p]
\centering
\includegraphics[width=0.96\textwidth,height=0.72\textheight,keepaspectratio]
  {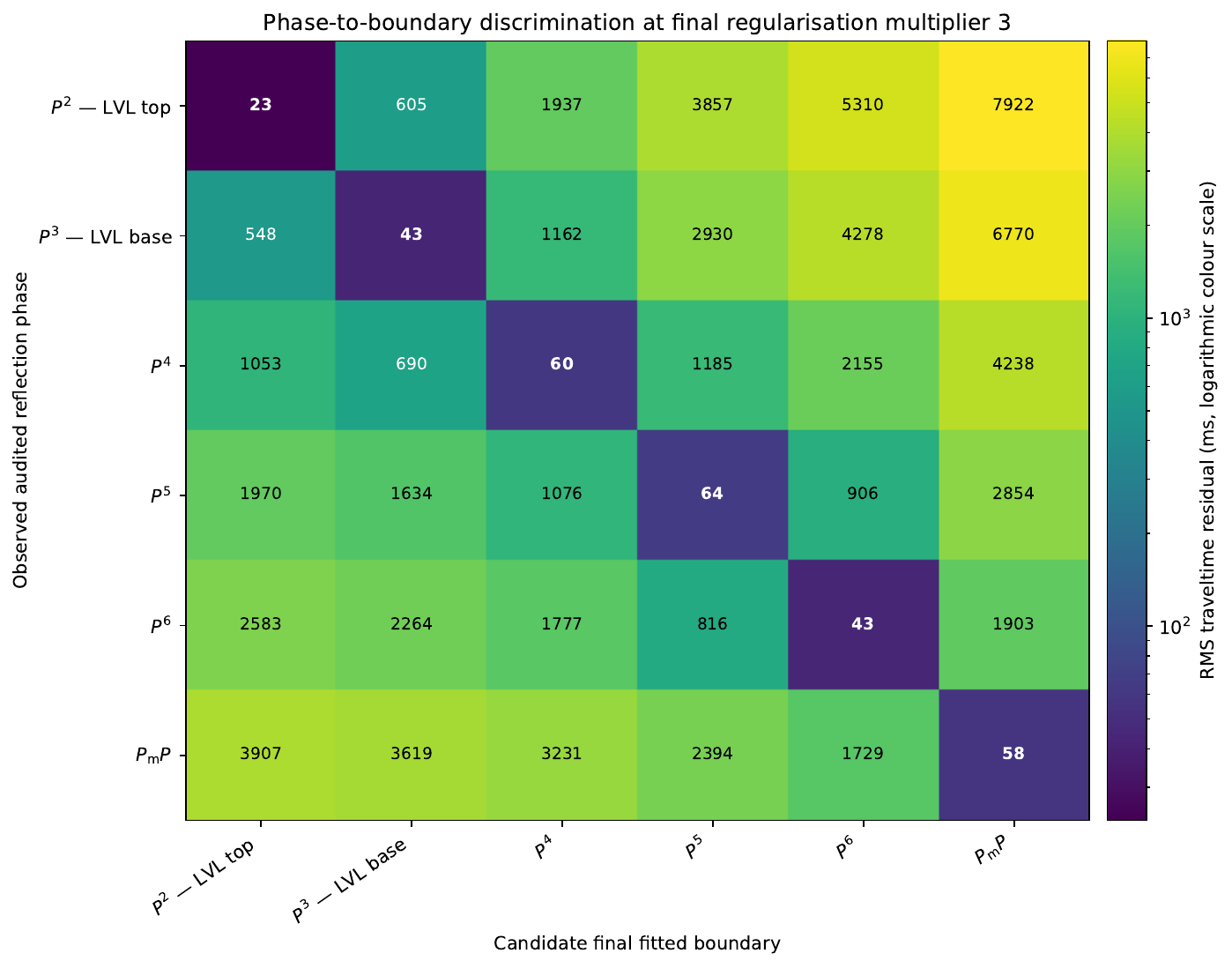}
\caption{Phase-to-boundary discrimination for the final fitted six-reflection
model at the final regularisation multiplier 3. Each audited reflection is
evaluated against every fitted candidate boundary; the correct diagonal is
the unique RMS minimum in every row. For \(P^{2}\) (ray 2.2, LVL top), the
diagonal RMS is 23~ms and the nearest wrong boundary, the \(P^{3}\) boundary,
gives 605~ms. For \(P^{3}\) (ray 3.2, LVL base), the diagonal RMS is 43~ms and
the nearest wrong boundary, the \(P^{2}\) boundary, gives 548~ms. Thus the
top- and bottom-of-LVL reflections are directly resolved as different
boundaries and are not interchangeable labels.}
\label{fig:supp-boundary-discrimination}
\end{figure}
\clearpage
\begin{figure}[p]
\centering
\includegraphics[width=0.99\textwidth,height=0.70\textheight,keepaspectratio]
  {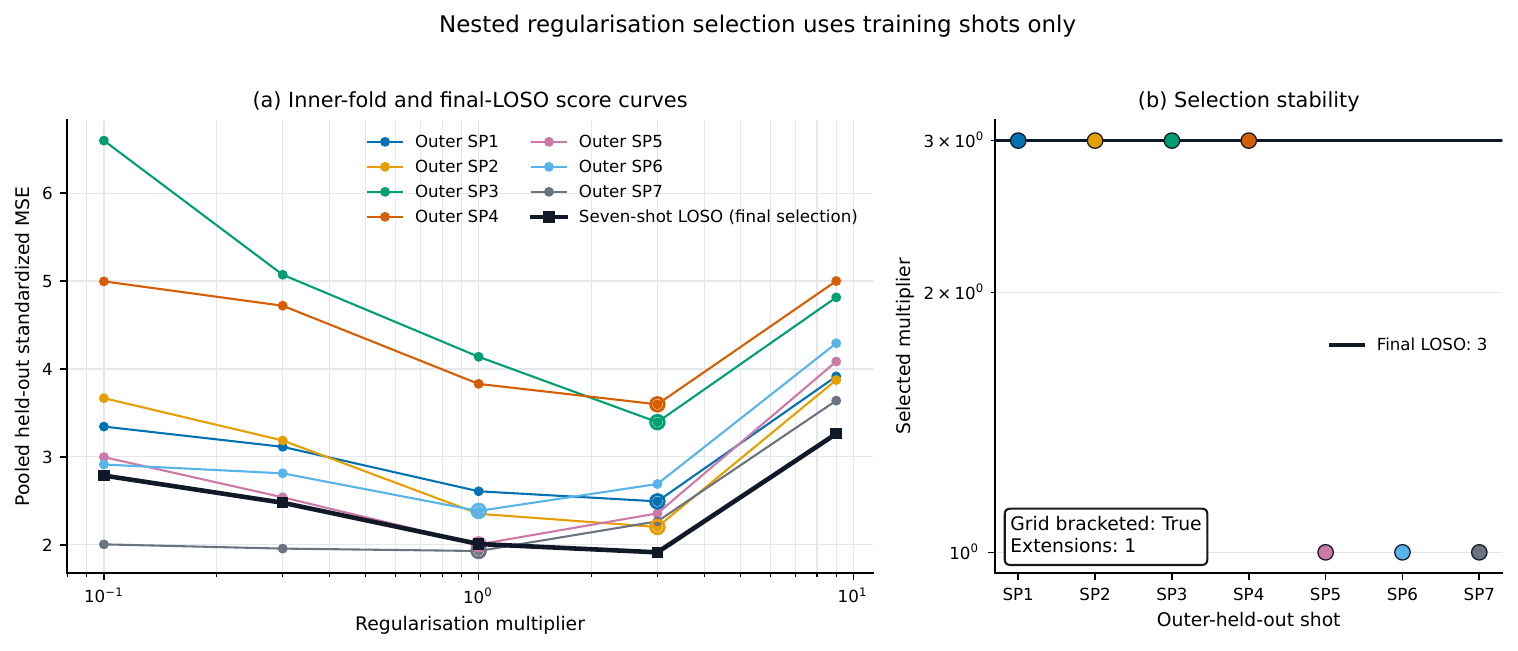}
\caption{Reflection regularisation selected only from training shots. The
extended evaluated multiplier grid is \(0.1,0.3,1,3,\) and \(9\). The
seven-shot full-data leave-one-complete-shot-out pass selects multiplier 3;
nested outer SP1--SP4 also select 3, whereas outer SP5--SP7 select 1. One grid
extension was used and no selected optimum lies on a boundary, so every
selection is bracketed.}
\label{fig:supp-reflection-lambda}
\end{figure}
\clearpage
\begin{figure}[p]
\centering
\includegraphics[width=0.99\textwidth,height=0.70\textheight,keepaspectratio]
  {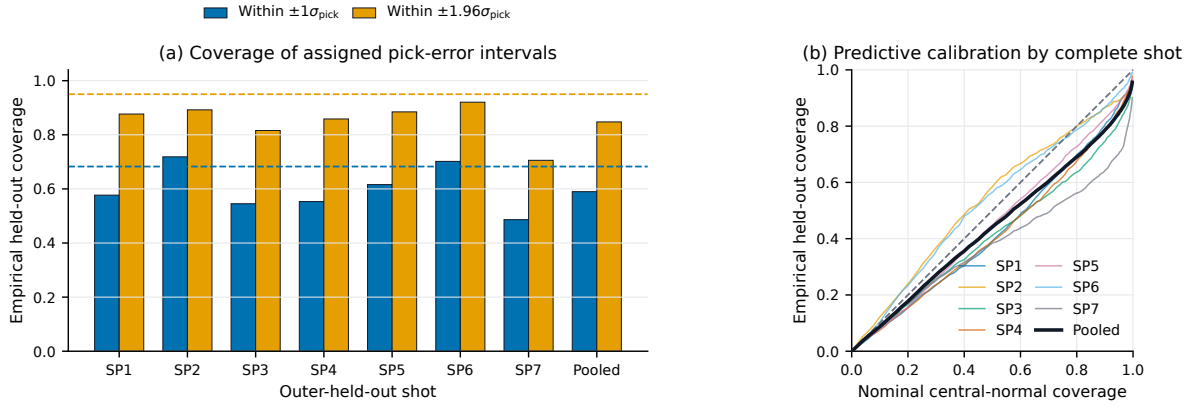}
\caption{Shot-wise nominal-interval coverage and central predictive
calibration for all 9,919 untouched primary-reflection predictions. Relative
to the pick-file uncertainties (25~ms for \(P^{2}\), 50~ms for
\(P^{3}\)--\(P^{6}\) and PmP), pooled empirical coverage is 0.590 within one
assigned standard deviation and 0.848 within 1.96 assigned standard
deviations. Because only seven complete shot clusters exist, these are
descriptive pick-level coverages conditional on the acquisition and model
class, not a distribution-free cluster-conformal guarantee.}
\label{fig:supp-reflection-calibration}
\end{figure}
\clearpage
\begin{figure}[p]
\centering
\includegraphics[width=0.96\textwidth,height=0.72\textheight,keepaspectratio]
  {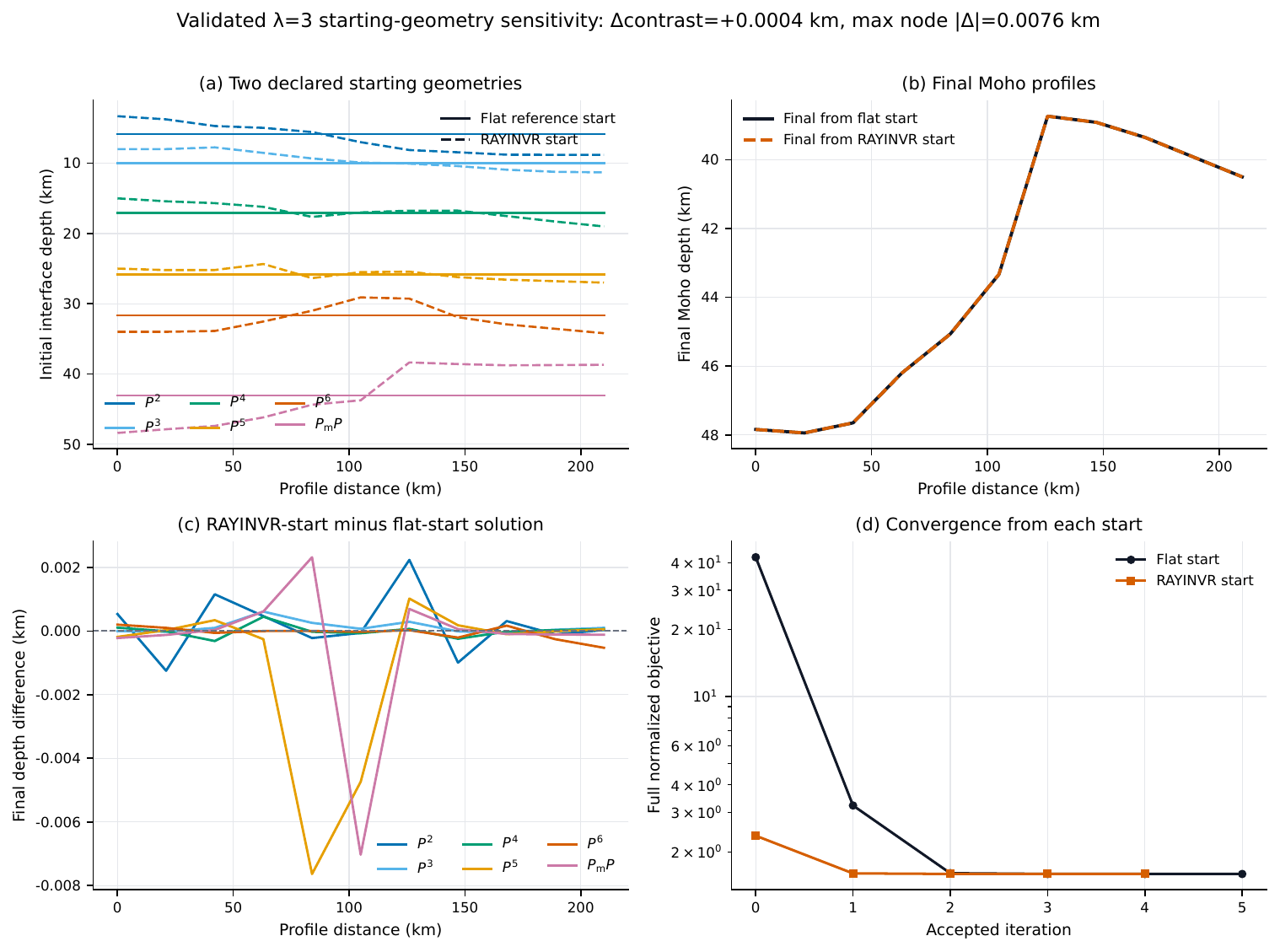}
\caption{Final fitted six-interface starting-geometry sensitivity at
regularisation multiplier 3. The solution started from the legacy
\textsc{rayinvr} interfaces minus the solution started from the flat reference
changes the exact WDC--CDC Moho contrast by only \(+0.00042594\)~km; the
maximum absolute difference across all fitted interface nodes is
0.007637~km. This comparison tests only these two declared starting
geometries. It does not test velocity, regularisation, parameterisation, or a
broader structural model class.}
\label{fig:supp-starting-geometry}
\end{figure}
\clearpage
\begin{figure}[p]
\centering
\includegraphics[width=0.98\textwidth,height=0.72\textheight,keepaspectratio]
  {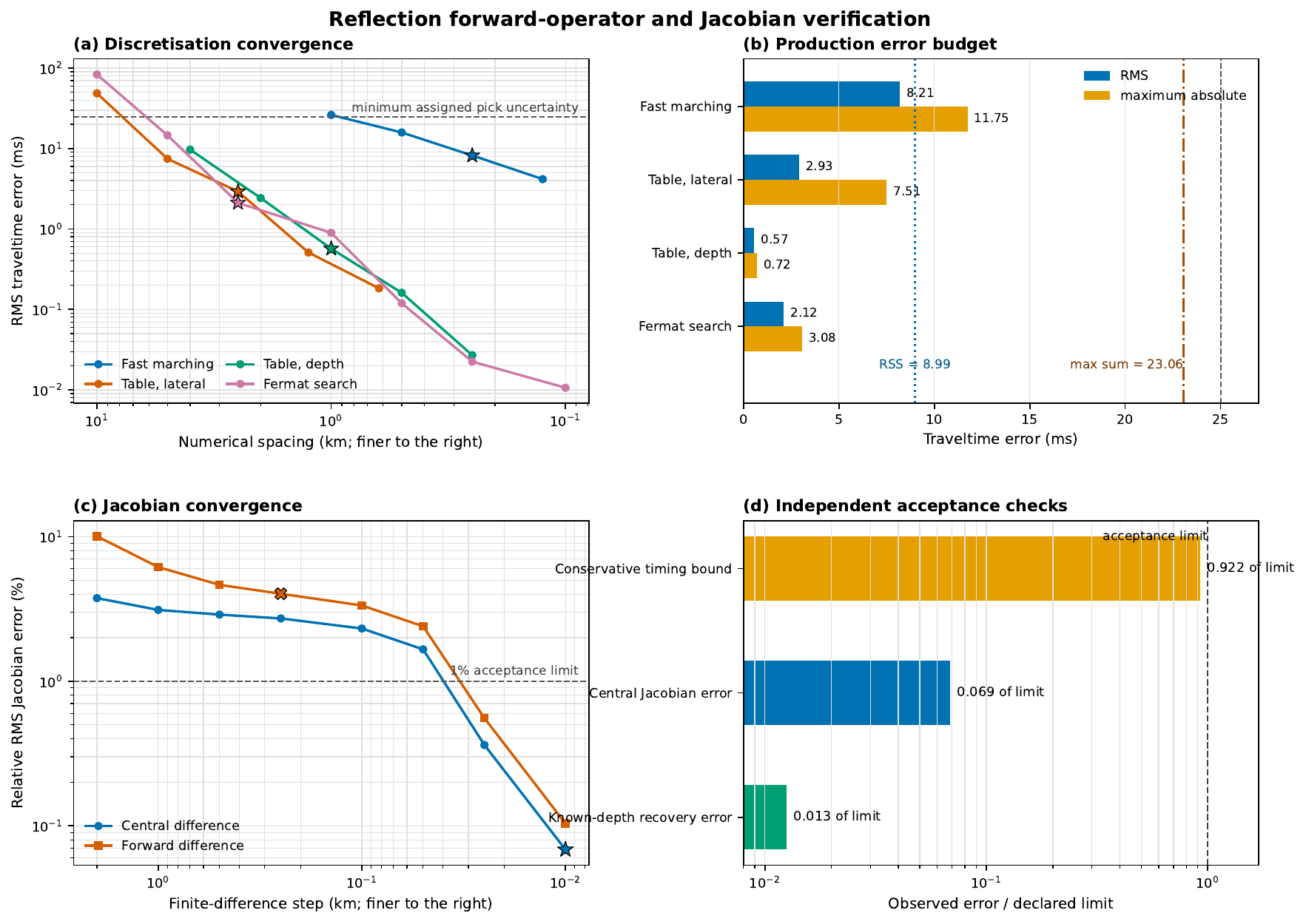}
\caption{Reflection forward-operator and Jacobian verification. At production
settings, the descriptive root-sum-square timing error is 8.986~ms and the
conservative sum of component maxima is 23.057~ms, below the minimum assigned
25-ms pick uncertainty. The central 0.01-km finite-difference Jacobian step has
0.0685\% relative RMS error. A noise-free flat reflector imposed at
22.350~km is recovered at 22.34875~km (depth error \(-1.252\)~m). The
conservative timing bound is a numerical verification limit, not an
independent random uncertainty component to add in quadrature.}
\label{fig:supp-reflection-numerics}
\end{figure}
\clearpage
\begin{figure}[p]
\centering
\includegraphics[width=0.99\textwidth,height=0.70\textheight,keepaspectratio]
  {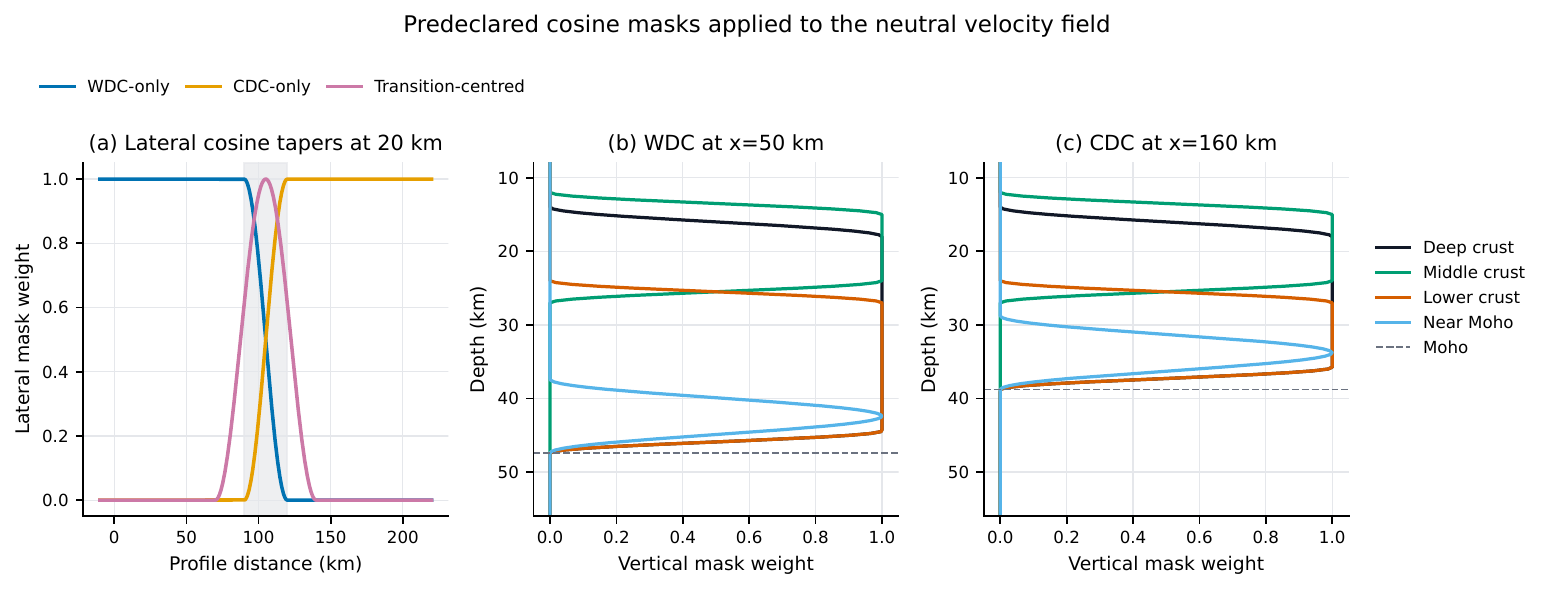}
\caption{Predeclared lateral and vertical cosine masks used in the full
nonlinear velocity--depth experiment: WDC-only, CDC-only,
transition-centred, deep-, middle-, lower-, and near-Moho perturbations. The
publication estimands average the linearly interpolated Moho exactly over WDC
\([0,90]\)~km and CDC \([120,210]\)~km. Across the intervening
90--120-km transition, the WDC and CDC lateral tapers are complementary.}
\label{fig:supp-velocity-masks}
\end{figure}
\clearpage
\begin{figure}[p]
\centering
\includegraphics[width=0.96\textwidth,height=0.72\textheight,keepaspectratio]
  {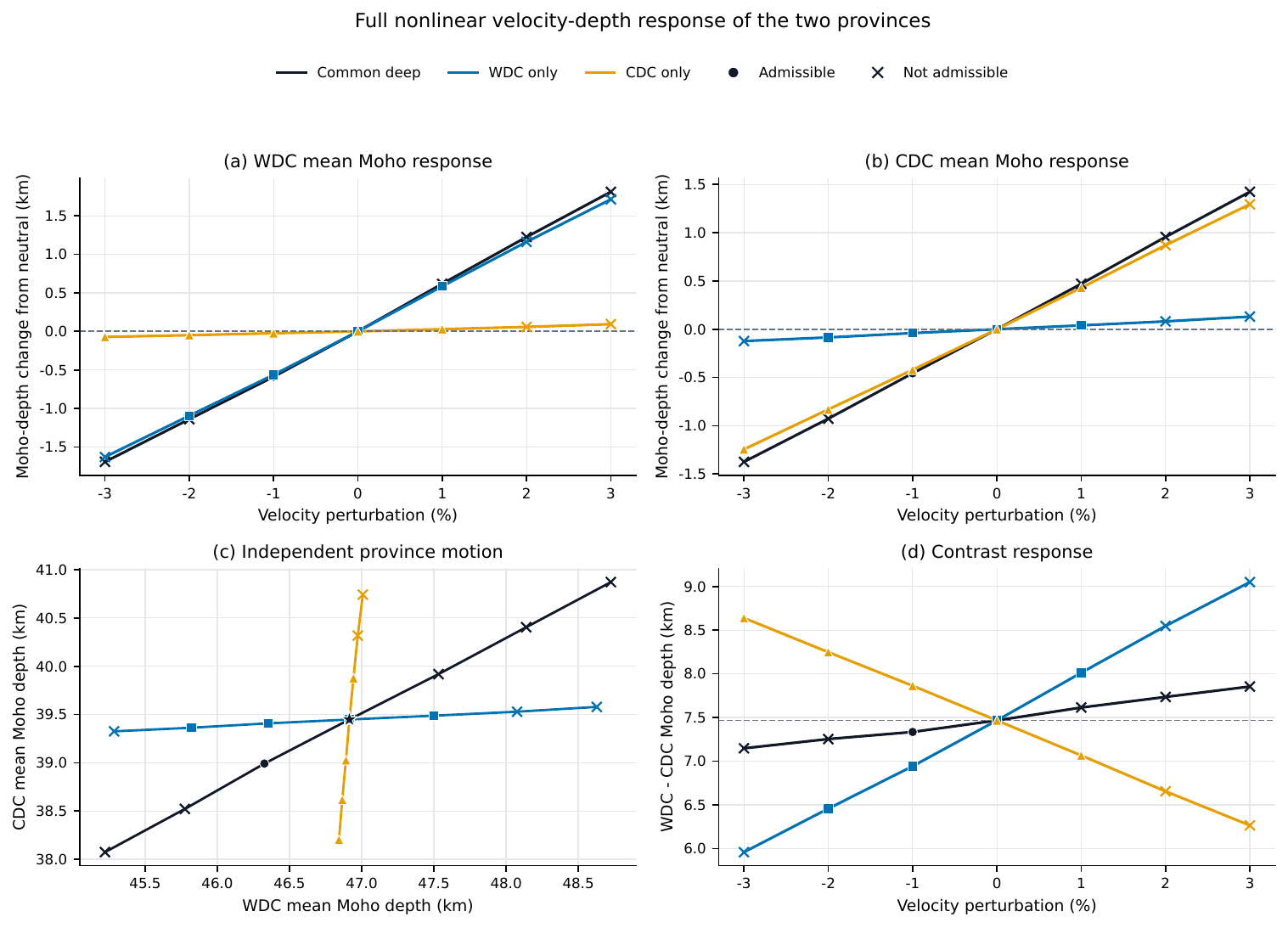}
\caption{Full nonlinear WDC and CDC Moho response to common and independent
province-specific velocity perturbations. Every unique perturbation rebuilds
the fast-marching reflection table and refits all six interfaces from the same
reference and start. Filled symbols pass the declared reflection,
interface-ordering, convergence, and applicable refraction gates; crosses do
not. Common deep perturbations move the two province means together, whereas
independent WDC-only and CDC-only perturbations change their difference. This
is why a single common velocity scaling is insufficient to characterize the
admissible Moho-contrast range.}
\label{fig:supp-province-response}
\end{figure}
\clearpage
\begin{figure}[p]
\centering
\includegraphics[width=0.96\textwidth,height=0.72\textheight,keepaspectratio]
  {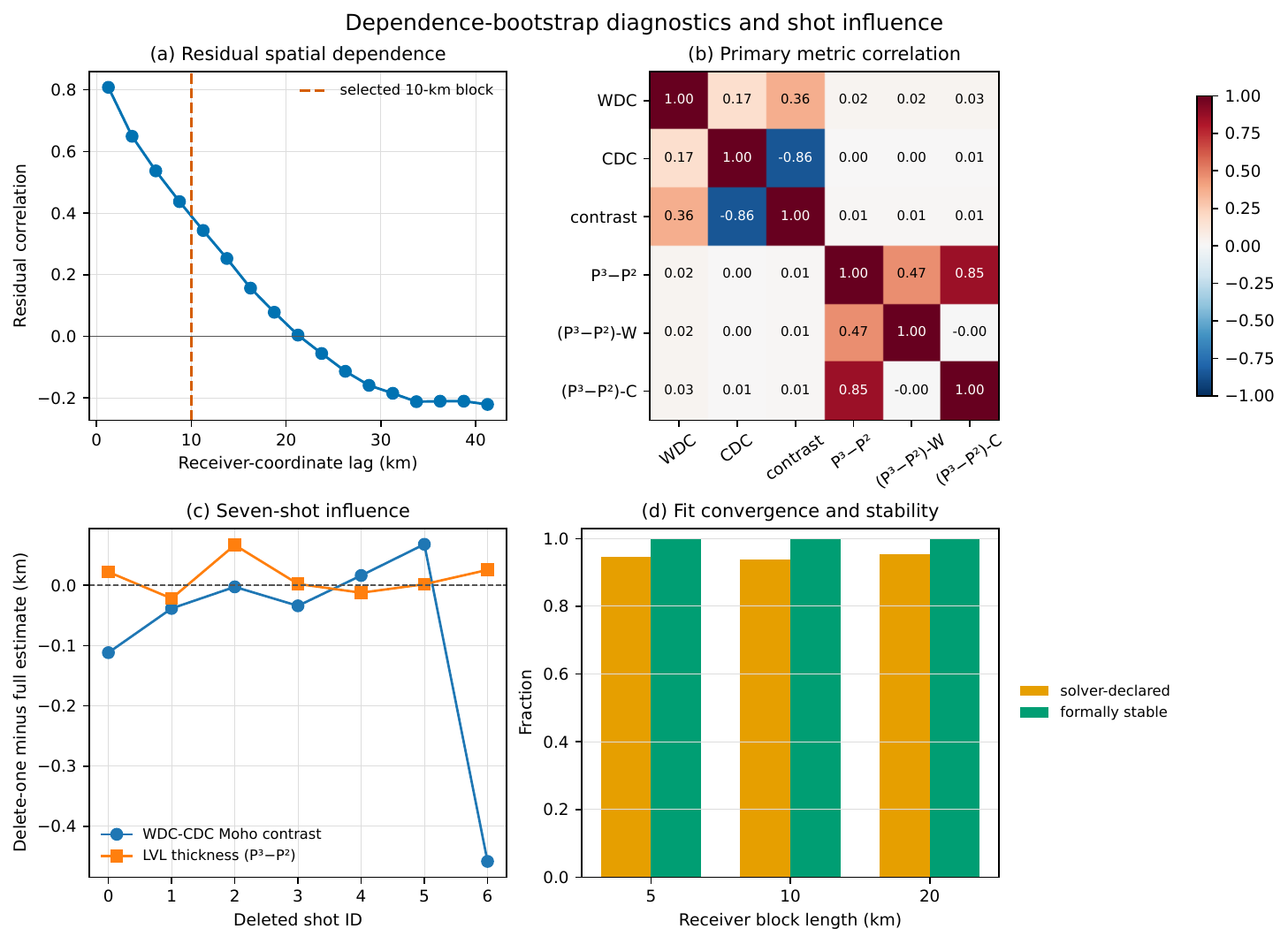}
\caption{Dependence-analysis diagnostics: residual spatial correlation,
bootstrap metric correlation, delete-one-complete-shot influence, and solver
stability. The estimated integral residual-correlation scale is 9.416~km, so
the nearest tested 10-km receiver-coordinate block is primary. Its 500
replicates preserve all phase picks and multiplicities at sampled receiver
coordinates. Delete-one-shot WDC--CDC Moho contrasts range from 7.008 to
7.534~km. With only seven shot clusters, this range is an influence diagnostic,
not a calibrated confidence interval; the receiver-block result is likewise
conditional on the fixed velocity, operator, regularisation family, and
observed shots.}
\label{fig:supp-dependence-diagnostics}
\end{figure}
\clearpage
\begin{figure}[p]
\centering
\includegraphics[width=0.99\textwidth,height=0.70\textheight,keepaspectratio]
  {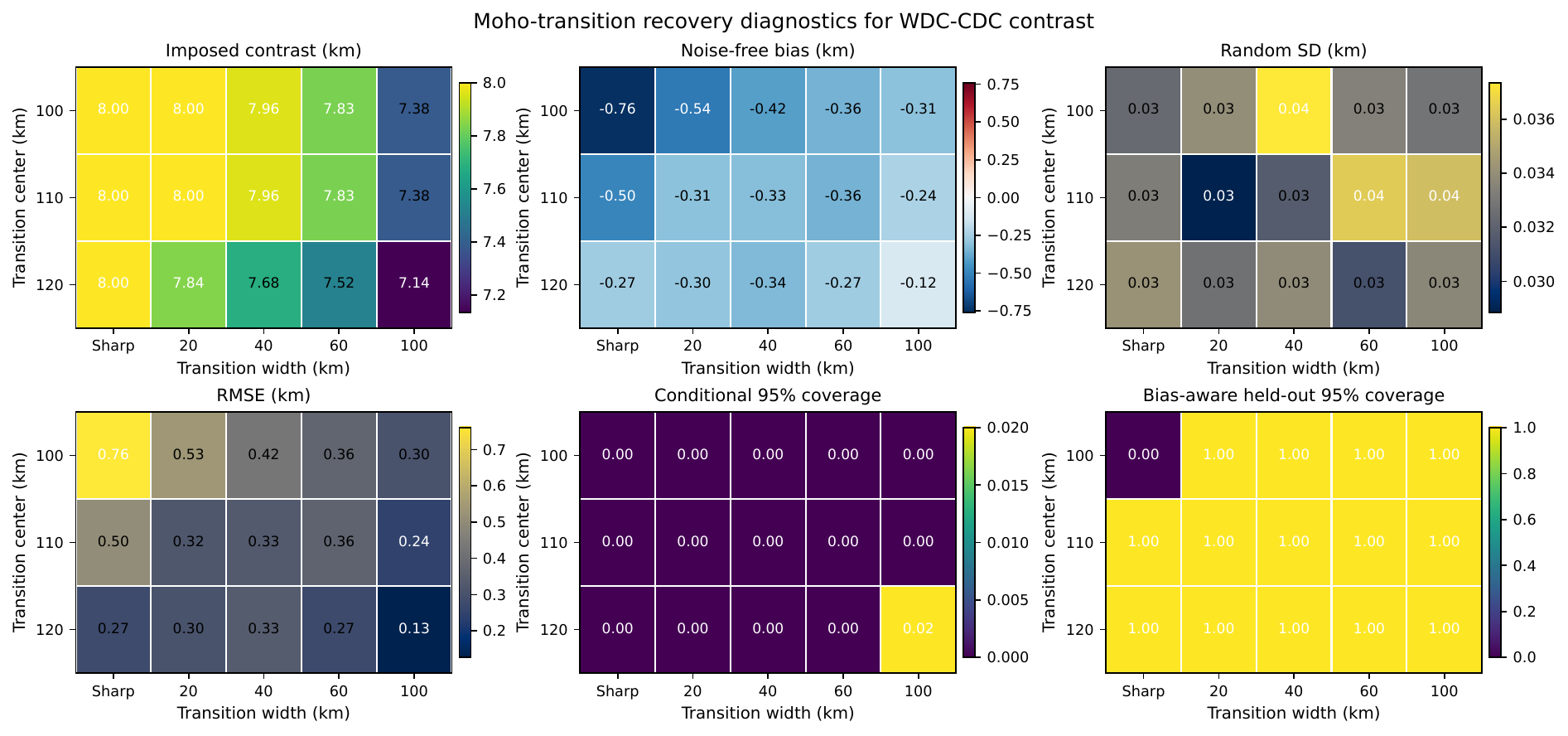}
\caption{Targeted Moho-transition scenario diagnostics for the WDC--CDC
contrast. Fifteen transitions combine centres at 100, 110, and 120~km with
sharp, 20-, 40-, 60-, and 100-km-wide forms; each has one noise-free fit and
50 independent pick-noise realizations, giving 750 noisy inversions. Panels
show imposed contrast, deterministic recovery bias, random standard
deviation, RMSE, uncorrected conditional coverage, and bias-aware held-out
coverage. Across scenarios, deterministic bias spans \(-0.760\) to
\(-0.124\)~km, whereas random SD is only 0.0289--0.0374~km, so recovery bias
dominates independent pick noise. The calibration is limited to the tested
transition family, actual PmP geometry, independent pick-file Gaussian noise,
and fixed 2-D velocity/operator class; it is not total uncertainty.}
\label{fig:supp-synthetic-diagnostics}
\end{figure}
\clearpage
\begin{figure}[p]
\centering
\includegraphics[width=0.99\textwidth,height=0.70\textheight,keepaspectratio]
  {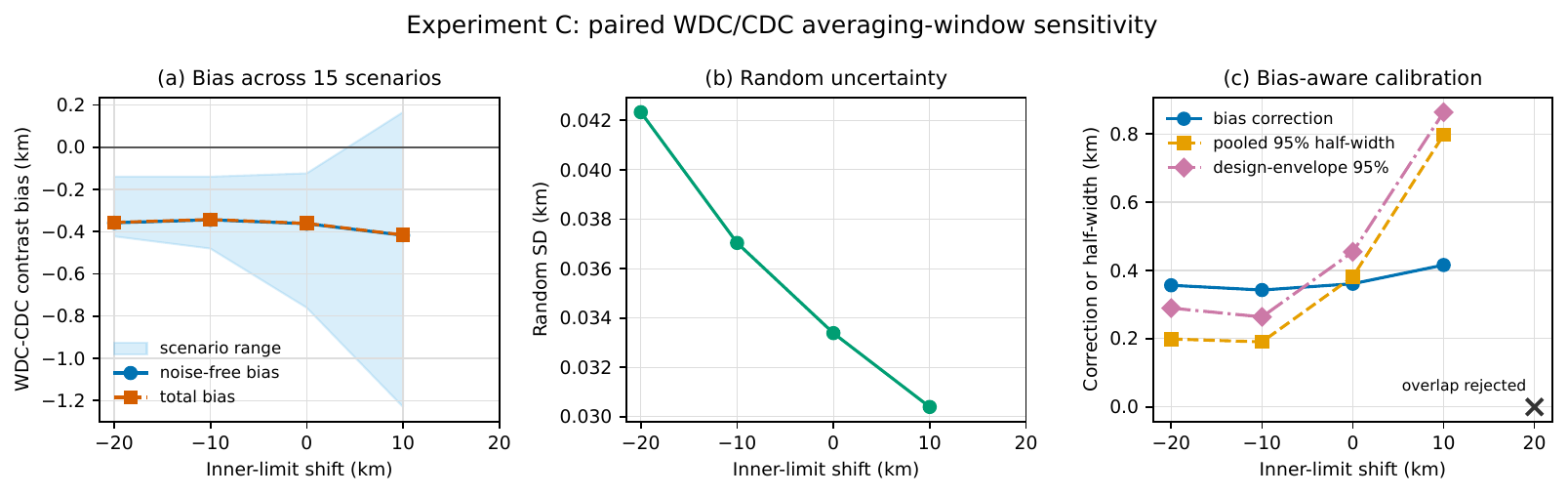}
\caption{Province-window sensitivity obtained by re-averaging every stored
v3 synthetic profile over four accepted, non-overlapping WDC/CDC window pairs:
outward 20~km, \([0,70]/[140,210]\)~km; outward 10~km,
\([0,80]/[130,210]\)~km; nominal, \([0,90]/[120,210]\)~km; and inward 10~km,
\([0,100]/[110,210]\)~km. The inward-20-km proposal,
\([0,110]/[100,210]\)~km, is rejected because the province windows overlap.
Panels compare recovery-bias ranges, random uncertainty, and bias-aware
calibration widths. These are re-averages of stored profiles; no new forward
or inverse runs are represented.}
\label{fig:supp-window-sensitivity}
\end{figure}

\clearpage
\section*{Supplementary table}

\begin{table}[H]
\centering
\small
\caption{Literature comparison of crustal thicknesses synthesized by
\citet{kumar2022thesis}. Values are in kilometres. ``NR'' means that
uncertainty was not reported in the thesis synthesis; no uncertainty is
inferred or invented here.}
\label{tab:supp-literature}
\begin{tabular}{@{}L{0.24\linewidth}L{0.30\linewidth}R{0.08\linewidth}R{0.08\linewidth}L{0.16\linewidth}@{}}
\toprule
\textbf{Study/source} & \textbf{Method in synthesis} &
\textbf{WDC} & \textbf{EDC} & \textbf{Uncertainty} \\
\midrule
\citeauthor{rao2015b} (2015b) &
Wide-angle/conventional seismic & 42 & 38 & NR \\
\citet{mall2012} &
Seismic layered model & 45 & 39 & NR \\
\citet{julia2009} &
Receiver functions and Rayleigh waves & 48 & 38 & NR \\
\citet{kumar2022thesis}; \citet{beherakumar2022} &
Current-profile legacy ray tracing & 49 & 38 & NR \\
\bottomrule
\end{tabular}
\caption*{\emph{Comparability warning.} WDC/EDC terminology follows the thesis
synthesis. Survey footprints, sampling, processing, and operational Moho
definitions differ among studies, so these rows are contextual rather than
matched estimates and must not be combined quantitatively with the present
exact-window WDC--CDC estimand.}
\end{table}

\end{document}